\documentclass[conference]{IEEEtran}
\usepackage[hyphens]{url}
\usepackage{xcolor}
\usepackage{graphicx}
\graphicspath{ {./images/} }
\usepackage{subcaption}
\usepackage{amsmath}
\usepackage{gensymb}

\usepackage[hidelinks]{hyperref}
\usepackage[lined,boxed,commentsnumbered]{algorithm2e}

\ifCLASSINFOpdf
\else
\fi
\begin{document}
\title{Orientation Estimation using Wireless Device Radiation Patterns}

\author{\IEEEauthorblockN{Thomas Burton}
\IEEEauthorblockA{Department of Computer Science\\
University of Oxford\\
Oxford, UK\\
Email: thomas.burton@cs.ox.ac.uk}
\and
\IEEEauthorblockN{Kasper Rasmussen}
\IEEEauthorblockA{Department of Computer Science\\
University of Oxford\\
Oxford, UK}}


%Telephone: (800) 555--1212, Fax: (888) 555--1212}


\maketitle

\begin{abstract}
%\boldmath
Wireless devices inherently have a non-uniform distribution of energy from their antenna or antennas. The shape that this forms is commonly called a \textit{radiation pattern} or \textit{antenna pattern}.
We demonstrate that orientation can be estimated without the cooperation of the target device despite only having a small number of RSS measurements per packet. We do this by applying bounds to the amount of rotation in the time interval between packets. Using simulations, we show that this method can achieve a mean orientation error as low as 7.6$\degree$. We then perform a security analysis to demonstrate the method's resistance to spoofing.
This paper focuses on consumer wireless devices where patterns are not deliberately highly directional and scenarios that cannot rely on contrived movement patterns of the entities involved, which is unrealistic or impractical in many settings. Our work concentrates on existing wireless systems and infrastructure common in domestic, office, and commercial.

\end{abstract}


% \fi
\IEEEpeerreviewmaketitle

\section{Introduction}

Localisation is an essential part of many modern systems. Global navigation satellite system (GNSS) and cellular localisation systems have allowed the creation of personal navigation systems, location services to personalise app experiences, emergency location beacons, and location-based multi-factor authentication systems to name just a few. Over the last decade, the popularity and number of indoor navigation applications have increased with the proliferation of smartphones and cheap sensor network hardware. Existing GNSS and cellular systems lack the accuracy required for personal navigation when indoors, so when it comes to indoor localisation and navigation, WiFi and Bluetooth received signal strength (RSS) \cite{bahl2000,mistry2015,sun2017,alabadleh2018,gui2017,hindmarch2016,ergen2014} and ultrasonic beacon systems \cite{xu2013,peng2007} have become favourites. One key problem with using RSS is that the power output by wireless devices is not uniform in all directions. If this inherent phenomenon is exploited, it potentially gives an attacker the ability to spoof their location fairly easily by using this different power output in different directions without the practical complexities of performing attacks using multiple transmitters or beamforming. These existing localisation systems also lack the ability to determine the orientation, which would open the door to a huge range of possible uses relating to the orientation of the device.

RSS, or \textit{RSSI}, is popular due to its wide availability on IEEE 802.11 compliant WiFi chipsets. RSS localisation schemes exploit the significant reduction in wireless transmission power over distance to provide estimates for range and thus location, or by using RSS fingerprints for locations. When it comes to measuring received power, several factors affect the measured value. One of the most interesting factors is antenna \textit{radiation patterns}. Ideally, mobile devices with a static radiation pattern, such as smartphones, would transmit energy equally in all directions so that the signal strength would be consistent no matter the device orientation. However, antennas are inherently more efficient at radiating power in certain directions, leading to an unequal distribution of energy in different directions from a transmitter. In addition, device components packaged around the onboard antenna will impact the energy distribution. Our experiments discussed later in this paper show that smartphones exhibit radiation patterns with significant differences in directionality in at least one plane. For example, for a Samsung Galaxy A50 we found it to have a maximum directivity of 5.1dBi and a minimum directivity of -12.5dBi. As a result, radiation patterns are responsible for a significant proportion of the positioning error in localisation \cite{mwila2014} and provide a rich source of additional information to exploit.

Initial approaches to deal with the problem of radiation patterns from a functional perspective discuss using onboard compasses \cite{coca2013,mwila2014}, this approach can provide corrections when performing self-localisation or when working cooperatively with infrastructure. However, as is the case with all data fusion approaches \cite{liu2014,kozlowski2018,yu2018,ying2019,redzic2020}, the required information is not always available. It may be undesirable to rely on information from a device under the control of an external entity, particularly in a security context where an adversary may spoof the onboard data. There may also be privacy concerns when sharing data of this nature. One example of this is work by Mwila et al. \cite{mwila2014} where they take a Gauss-Newton optimisation approach to find the location while correcting for the radiation pattern cooperatively. This provides very efficient estimates for location and a significant improvement on methods that do not consider radiation patterns. It does not consider that due to the nature of radiation patterns, it is possible to find multiple regions that with different orientations provide a very close match. This is mainly a problem with a small number of receivers (e.g., 3 or 4) or nodes in the case of sensor network configurations.
  
More recent work by Zuo et al. \cite{zuo2019} focused on localisation in a different setting, one in which the system must estimate specific parameters, such as transmission power, location, and orientation. However, like many other works in the wireless sensor networks (WSN) domain, this work assumes that there is a single main beam which may be appropriate in some applications but not in the smartphone domain. They also make the powerful assumption that the receiver nodes have isotropic patterns, and this potentially misses a rich source of information. However, in their case, the sheer number of receivers in their approach adds enough constraints that the probability of many different areas of strong matches is very low despite the flexibility provided by the radiation pattern.

The issues of radiation patterns have been explored extensively for wireless sensor networks (WSN) \cite{george2020}, where deliberately using a highly directional antenna is very advantageous in terms of reducing power consumption and RF interference. In these scenarios, it is often possible to model radiation patterns as a single Gaussian-shaped lobe or cone projecting outwards from the transmitter \cite{nasipuri2002,ou2011,jiang2012,chang2018,zuo2019,wielandner2021}. However, when considering mobile consumer devices, such as smartphones, simple models for the pattern are no longer appropriate as they often use an omnidirectional antenna whose pattern is significantly affected by other hardware components packaged into the device. As a result, the nature of the pattern is much more like the patterns measured by \cite{coca2013} and \cite{mwila2014}, and maybe an arbitrary shape. It should also be noted that for localisation in sensor networks, the steps to be carried out can, in some cases, be slightly contrived. For example, receivers may be arranged in a particular formation \cite{ou2011,jiang2012} or perform a specific type of motion \cite{chang2018}. While perfectly applicable in WSN contexts, schemes with these characteristics are harder to justify in commercial, domestic, or office environments where the localisation system uses existing and normally static networking infrastructure. 

There is also a particular type of cooperative localisation that is very popular in sensor networks \cite{mwila2014,wielandner2021} where multiple devices being localised cooperate, the idea is that by communicating with other nodes that are also trying to determine their location, some additional information can be obtained. This type of localisation does have clear applications outside of sensor networks, for example, localisation in crowds or disaster recovery scenarios, but this is very distinct from our work where there is a single device to localise.

Many existing schemes use multiple positions estimates over time to analyse the trajectory to make the overall trajectory more realistic. This can be done by either 1) constraining the position or movement using constrained paths \cite{ergen2014,liu2014,hindmarch2016} or a model of a devices movements \cite{bonne2013,liu2014}; or 2) using smoothing functions applied to measurements or the trajectory itself to produce more realistic trajectories from those with large amounts of jitter \cite{nordlund2002,sarkka2015,yu2018,luo2018,choi2022}.

Many pieces of related work leverage radiation patterns, and while their goals may be similar, the context and processes make them distinct from what we are trying to achieve. We aim to leverage radiation patterns in more conventional commercial, domestic, and office environments with more limitations on the infrastructure, available data, and a broader range of radiation pattern shapes.
We apply the idea of constraints to the device's rotation to smooth the trajectory. We propose using radiation patterns to estimate the orientation of a device remotely and non-cooperatively and classify orientation values as outliers if they show unrealistic changes in orientation. This work aims to improve upon the accuracy of existing RSS localisation schemes in a scenario where the infrastructure performs the localisation and does not have onboard sensor data. The advantages of this approach are that we can use existing wireless infrastructure that passively performs the localisation without the device's cooperation. RSS is particularly suited to this as it is available on almost all networking hardware due to its valuable network management and troubleshooting role.
Our approach means that we need continuous packet transmission to bound the rotation. However, this provides strong security guarantees as spoofing a path becomes much more difficult.

To the best of our knowledge, no one has exploited the additional information of radiation patterns from mobile devices to improve RSS localisation by filtering outliers based on improbable rotation.

In this paper, we make the following contributions:
\begin{itemize}
	\item We propose a novel localisation scheme that applies constraints to a device's rotation to eliminate possible locations and orientations that would require improbably fast rotation speeds to achieve. 
	
	\item The proposed scheme is designed to work with any arbitrarily shaped radiation pattern, such as those found in smartphones. The scheme can be integrated with existing COTS networking infrastructure and only relies on a small number of receivers.
	
	\item We demonstrate the feasibility of this by measuring the radiation patterns from several smartphones and use simulations to estimate the level of accuracy that can be achieved in terms of orientation and position.
	
	\item We perform a security analysis to demonstrate the scheme's resistance to path spoofing attacks.

\end{itemize}

The remainder of the paper is laid out as follows: In Section~\ref{sec:background} we cover the technical background necessary for understanding our work. In Section~\ref{sec:cLocalisation-systemModel} we define the system and attacker model, and in Section~\ref{sec:cLocalisation-concept} we explain the concept of using radiation patterns to perform localisation, the challenges, and our solutions. In Section~\ref{sec:cLocalisation-Design} we present the detailed design of our enhanced localisation pipeline. In Section~\ref{sec:cLocalisation-experiments} we evaluate the scheme and present the practical experiments and simulations that were performed to show the feasibility of the key elements of the pipeline. In Section~\ref{sec:cLocalisation-spoof} we perform a security analysis to demonstrate the resistance of the scheme to particular spoofing attacks.  In Section~\ref{sec:related} we discuss related work.
Finally, we conclude in Section~\ref{sec:cLocalisation-conclusion}.

\section{Technical Background}\label{sec:background}

We now present the required background knowledge required to understand the remainder of this paper.

Antennas do not radiate power equally in all directions and the resulting pattern of transmission energy is called a \textit{radiation pattern} (or \textit{antenna pattern}) \cite{balanis2016}. For a particular antenna, this can be represented as a mathematical function or graphically to show the radiation properties as a function of space coordinates. To consider where a particular point in space is relative to an antenna, we use the radius $r$ from the antenna, azimuth $\phi$, and elevation $\theta$. The most important property for this research is energy distribution in different directions.

As stated in \cite{balanis2016}, an isotropic radiator is a ``hypothetical lossless antenna having equal radiation in all directions''. However, this is not physically realisable, and is instead taken as a reference for discussing the directivity of real antennas. A directional antenna has ``the property of radiating or receiving electromagnetic waves more effectively in some directions than in others. This term is usually applied to an antenna whose maximum directivity is significantly greater than that of a half-wave dipole.'' When we consider that antennas have a 3-dimensional pattern, it may be the case that the pattern is directional in one plane, while in another, it is uniform. This is considered an omnidirectional pattern.

It is important to note that there is a change in the pattern of an antenna as the distance viewed moves through the different field regions: reactive near-field, radiating near-field (Fresnel), and far-field (Fraunhofer). In the closest region---reactive near-field---the pattern is more uniform, but lobes form in the further out regions. It is necessary to consider this in any applications of radiation patterns.

How radiation patterns manifest is primarily dependent on the type and construction of the antenna. Strong directional lobes can be achieved with specific types of antennas that radiate strongly in a particular direction, for example, with an aperture antenna. Additional directionality can be realised by exploiting constructive and destructive interference, either with an array of antennas or by using other antenna elements. Reflectors can also direct radiation in a particular direction, for example, a parabolic reflector. Even if unintentional, the construction of the device surrounding an antenna also affects how the pattern manifests. Therefore, we use the term \textit{radiation pattern} rather than \textit{antenna pattern} as we are referring to the pattern produced by the device as a whole, not just the antenna.

The principle of directionality has been used to great effect for many different applications, for example, in satellite communication, the power consumption is reduced by using a highly directional beam to target ground stations; in aviation, navigation systems make use of directional antennas and phased antenna arrays; and in mobile communications, there is a clear trend in the direction of beamforming both in 5G \cite{3gpp2019} and WiFi \cite{2013} networks where multiple antennas are used to carefully craft specific pattern shapes to control the direction of the signal to improve communication quality further, reduce power usage, and improved throughput using spatial multiplexing. In the field of sensor networks, there has been some security work leveraging radiation patterns. Solutions have been proposed to resolve common security issues with sensor networks like wormhole and Sybil attacks \cite{hu2004,george2020}.
The commonality of these methods is that they are deliberately using radiators that are specifically designed to be highly directional.

\subsection{Directivity}

\textit{Directivity} of an antenna is defined as ``the ratio of the radiation intensity in a given direction from the antenna to the radiation intensity averaged over all directions'' \cite{ieee2014}. ``The average radiation intensity is equal to the total power radiated by the antenna divided by 4$\pi$'' \cite{balanis2016}. In essence, directivity gives us the ratio compared to the theoretically perfect isotropic radiator, which has a directivity of 1 in all directions. If the direction from the antenna is not specified for a particular measurement, then it is typically assumed to be the direction of maximum radiation intensity. From the definition, the directivity for a given elevation and azimuth, $D(\theta,\phi)$, is calculated by: 
\begin{equation} \label{eq:lit-directivity}
	D(\theta,\phi) = \dfrac{ U(\theta,\phi) }{P_{rad} / (4 \pi)}
\end{equation}
Where $P_{rad}$ is the total radiated power output, and U is the radiation intensity at the specified angle. $P_{rad}$ may be found through a number of methods: using the chipset specification; through calculation using input power and efficiency; an estimate can be retrieved from the US Federal Communications Commission (FCC) certification test report; or by calculation of $U(\theta,\phi)$ integrated over the spherical surface if the pattern is represented by an equation. 
\begin{equation}
	P_{rad} = \int_{0}^{2\pi} \int_{0}^{\pi} U \sin \theta \,d\theta\, d\phi 
\end{equation}
Directivity has dimensionless units, but often it is expressed in decibels relative to a reference radiator. Using a logarithmic scale has the advantage of 
more strongly highlighting very low values, which is particularly useful when considering the minor lobes of a radiation pattern.

\subsection{Data Representation}

A radiation pattern can be represented as a collection of tuples consisting of azimuth, elevation, and directivity values. The data structure used to store these tuples may vary based on the particular application. Throughout this paper, we primarily represent patterns as a 2-dimensional array for simplicity and computational efficiency. The first dimension is the azimuth, the second is the elevation, and the value stored is the directivity. With the azimuth and elevation indexes having a constant step, this does create the effect of having higher measurement densities at some parts of the pattern.

\section{System and Attacker Model}\label{sec:cLocalisation-systemModel}

The localisation system has an area of operation $\mathcal{A}$. Transmitter \textit{T} connects to a wireless access point (\textit{AP}) which is part of the wireless infrastructure of a network. \textit{T} is a smartphone capable of bi-directional communication with the AP and uses this to access network services. For localisation, there are a set of $n$ receivers ($R_0$, ..., $R_n$) and a central server (CS) that performs the localisation calculations. The receivers are passive on the wireless network, and they communicate with CS using separate wired connections. The receivers are positioned around the environment at fixed locations and orientations known to CS, and this information is stored in \textit{receiver database}. The radiation patterns of enrolled devices may be of any arbitrary shape.
The patterns are stored in \textit{pattern database} so the radiation pattern of \textit{T} is known. All entities of the system have static radiation patterns.

Transmitter \textit{T} transmitts packets to communicate with the AP. Any receiver within range records the packets. The packet data is then sent to CS along with packet metadata, including sender and receiver MAC addresses, the approximate time the packet was received, and the RSS value. The CS then performs calculations to determine the location. The CS does not have access to data from onboard sensors from the device. 

As we will discuss, a site survey must be conducted, and this is performed using a transmitter with a known pattern.

\subsection{Attacker Model}\label{sec:attacker-model}

The proposed scheme has the following security guarantee: When given a path of $k$ waypoints consisting of positions and orientations, the attacker cannot find another valid path in $\mathcal{A}$ that will spoof the path defined by the waypoints.

We later discuss and analyse spoofing attacks on the system in Section~\ref{sec:cLocalisation-spoof}. In that analysis, we use the following attacker model:

An adversary attacks the scheme to spoof her location and orientation along a particular path. By which we mean to make it appear as if she passed through a set of $k$ waypoints at particular locations (i.e., specific locations within rooms or corridors), that she did not pass through and with the device appearing to be at a particular orientation. The set of waypoints of length $k$ is specified by the CS.

We make the following assumptions about her capabilities: 
\begin{enumerate}
	\item She must conduct her attack from within $\mathcal{A}$.
	\item She has a single antenna with a fixed radiation pattern that she may not modify during the attack (i.e., no beamforming or physical alteration between packets). Notably, the pattern has not been specifically designed to attack a particular configuration of receivers, but it may be different from the one the system believes is being used.
	\item She may adjust the energy output by increasing or decreasing the transmission power.
	\item The interaction between the attacker and the CS is configured in such a way that the CS believes that she is using a device with a radiation pattern that is in the \textit{pattern database}.
	\item She has complete knowledge of the scheme, including configuration data used by the CS, such as the site survey data and pattern database.
	\item She may not perform a denial of service attack on the whole system, the CS database, or the communication between the receivers and the CS.
	\item She may not read or modify the communications between the receivers and the CS.
	\item She may not modify the contents of the \textit{pattern} or \textit{receiver databases}.
	\item She may not rotate or move the transmitter any faster than their respective limits that we discuss later in the description of our localisation scheme.

\end{enumerate}

To make this attack meaningful, we define another condition: she cannot spoof her location by simply following the spoofed path as that would not be a meaningful attack. Therefore, she must be separated from the position to spoof by at least distance $d$.

\section{Localisation using Radiation Patterns}\label{sec:cLocalisation-concept}

\begin{figure*}[t]
	\centering
	\includegraphics[width=0.80\textwidth]{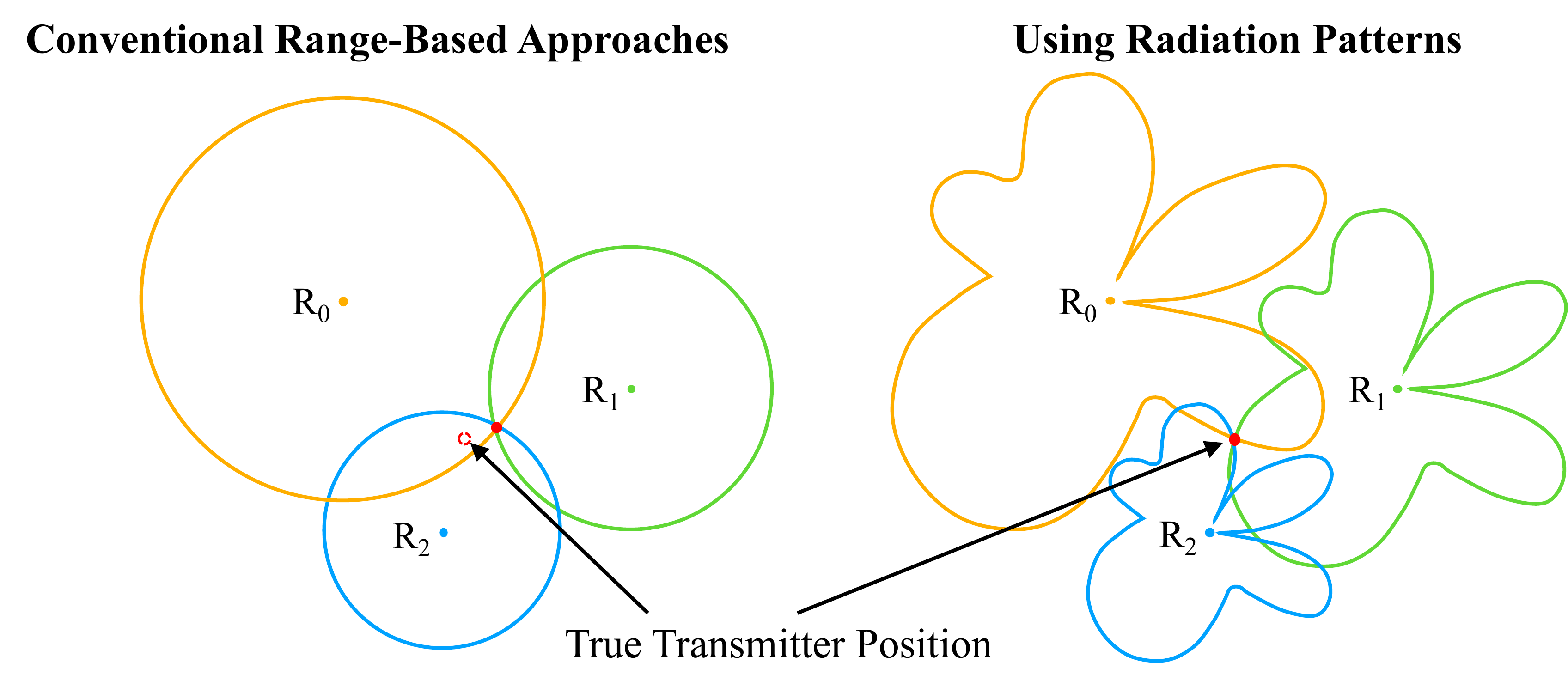}
	\caption[RSS Radiation Pattern Localisation]{Diagram showing conventional range-based approaches versus using the radiation pattern to estimate the location. The estimated position is marked by a red dot and in the conventional case the true position is marked by a dashed circle.}
	\label{fig:concept}
\end{figure*}

\begin{figure*}[t]
	\centering
	\includegraphics[width=1\textwidth]{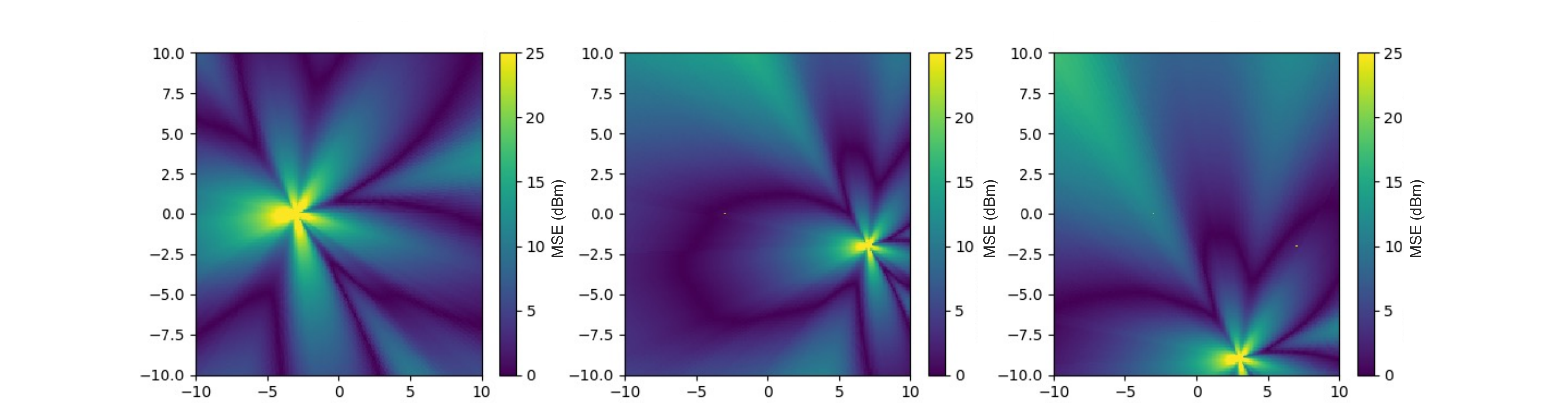}
	\caption[RSS Radiation Pattern Localisation]{Diagram showing maps displaying the mean squared error at different points relative to three receivers. The areas where the dark areas overlap for the three receivers represent a strong position match, and the rotation of the pattern for the strong match represent a strong match for orientation. }
	\label{fig:concept-power}
\end{figure*}

Radiation patterns manifest as an irregular distribution of energy from a transmitting antenna. This provides additional information that can be exploited to enhance localisation by considering the possible orientations of the device. 
Rather than considering RSS as just a function of path loss, transmission power, and slow fading, we model the measured RSS as a combination of these plus the impact of the radiation patterns of the transmitter and receiver.

Intuitively, we can see that a model that considers antennas isotropic or omnidirectional will have some limitations when the actual pattern has significantly higher directivity in some directions. Although we will later use a location fingerprinting approach as our solution, we will first illustrate the problem with a range-based approach as this is simpler to understand. We illustrate this with the 2D example shown in Figure~\ref{fig:concept}, where there is a single transmitter (its estimated position shown in red) with a known radiation pattern and three receivers (shown in orange, green, and blue). With conventional range-based approaches, the goal is to find the closest point of intersection of the three circles. However, with radiation patterns, we need to consider both the distance from each receiver and the relative orientation of the various devices. For simplicity, in this example, we will assume the receivers are omnidirectional in the measurement plane, so we only consider the transmitter's pattern. On the right-hand diagram of Figure~\ref{fig:concept}, each line marks all the positions at which the received power would be constant for that receiver.
For each receiver there is an RSS measurement, $\left[rss_0, rss_1, rss_2\right]$. The values are combined with the known pattern to calculate the distance in different directions from the receivers that correspond to the measured RSS values. Each coloured line represents the distance for a particular measurement, i.e. any position on the line will give the same RSS value. The patterns are then rotated through the possible orientation state-space to find the closest point of intersection. 
The closest point of intersection is taken as the estimated position $(x,y)$, and the rotation necessary to achieve this intersection is the orientation $(\phi)$. This can also be viewed as a map showing the mean squared error for each receiver, as shown in Figure~\ref{fig:concept-power} which shows a similar receiver layout.

\subsection{Challenges}

In practice, we are presented with the following challenges that make this method difficult to perform:

\begin{enumerate}
	\item[C1] \textit{Limited receivers.} The number of receivers is limited, and a receiver can only measure the part of the pattern that is in its direction.
Only a small number of points influenced by the pattern can be measured for one packet. This may make it hard to narrow down the possible positions and orientations to one. As a result, there will potentially be multiple arcs where the observed patterns cross, resulting in multiple highly probable locations.
	
	\item[C2] \textit{Patterns with regions of symmetry.} Some patterns may exhibit some elements of symmetry, whether that is in a single plane or one section of the pattern. This further adds to the previous challenge.
	
	\item[C3] \textit{Measurement noise.} For a set of samples from a single packet, there will be some level of measurement noise, and as with all localisation schemes this will lead to some inaccuracy. 
	
	\item[C4] \textit{Closed form radiation pattern equations.} Ideally, a closed-form 3D equation for each antenna pattern would be used, and a system of equations could be used to solve for position and orientation. However, generating the equation for the 3D radiation pattern may sometimes be very challenging to do automatically. The number of required coefficients can vary greatly, and if not carefully monitored, important details of the pattern can be lost.
	
	\item[C5] \textit{Noise reduction and power-saving techniques.} Some wireless devices use techniques to reduce noise or power consumption. Beamforming uses constructive and destructive interference from multiple antennas to target the device's location. This changes the shape of the radiation pattern. While adaptive transmission power techniques increase or decrease the transmission power to find an optimal balance of energy consumption and noise, this does not alter the pattern shape.
	
	\item[C6] \textit{Device Heterogeneity} When collecting samples from the site survey, the transmitter being used has a pattern that will influence the RSS measurements taken. If not corrected, the site survey data will not be suitable for our purposes.

\end{enumerate}

\subsection{Solutions}

Firstly, we resolve C1 by combining movement with multiple packets over time. As a device moves through an environment with a limited number of static receivers, different parts of the radiation pattern will likely be exposed to the receivers, and more extensive pattern coverage will be achieved with a small number of receivers. Combining movement and multiple packets also goes some way to resolving C2 and C3. However, to fully resolve them, we need to apply bounds to the device's rotation, which adds constraints that can eliminate some possible incorrect positions and orientations. 
To apply reasonable bounds on the device's rotation, we need continuous packet transmission.

To resolve C4, patterns are stored as 2D slices of the pattern at different elevations with discrete values for different azimuths. Next, because it is difficult to use a system of close form equations to find the location and orientation, we need to compare the expected RSS values from the different possible candidate locations and orientations to the actual RSS values measured, as is normal with location fingerprinting approaches.

Fortunately, when it comes to C5, beamforming is not standard in smartphones as the current and previous generations have a single WiFi antenna. However, variable transmission power may be used on smartphones. On the devices we investigated, this was not an issue. If it were to be used, it would be possible to add transmission power as another unknown candidate variable, and this does not prevent the use of radiation patterns for orientation estimation.

To resolve C6, the data collected from the site survey must have the transmission power and directivity of the transmitter removed. What is stored for later use is a combination of path loss, slow fading, and the receiver's directivity.

Putting these solutions together, we propose a scheme that can be summarised as follows: RSS samples are taken by $n$ receivers as $T$ moves through the environment. Different candidate locations and orientations for each packet are compared to RSS samples collected during a site survey. A large number of the closest matches are clustered into groups of location and orientation using a clustering algorithm. Each of the clusters is compared to the previous valid location and orientation. The cluster that is not valid with respect to a model that bounds the rotation and then movement is eliminated. The centroid of the clusters of the closest remaining matches is then accepted as the valid location and orientation for the packet. The process is then repeated for the next packet.

\section{Enhanced Localisation Pipeline} \label{sec:cLocalisation-Design}

\begin{figure}[t]
	\centering
	\includegraphics[width=1\columnwidth]{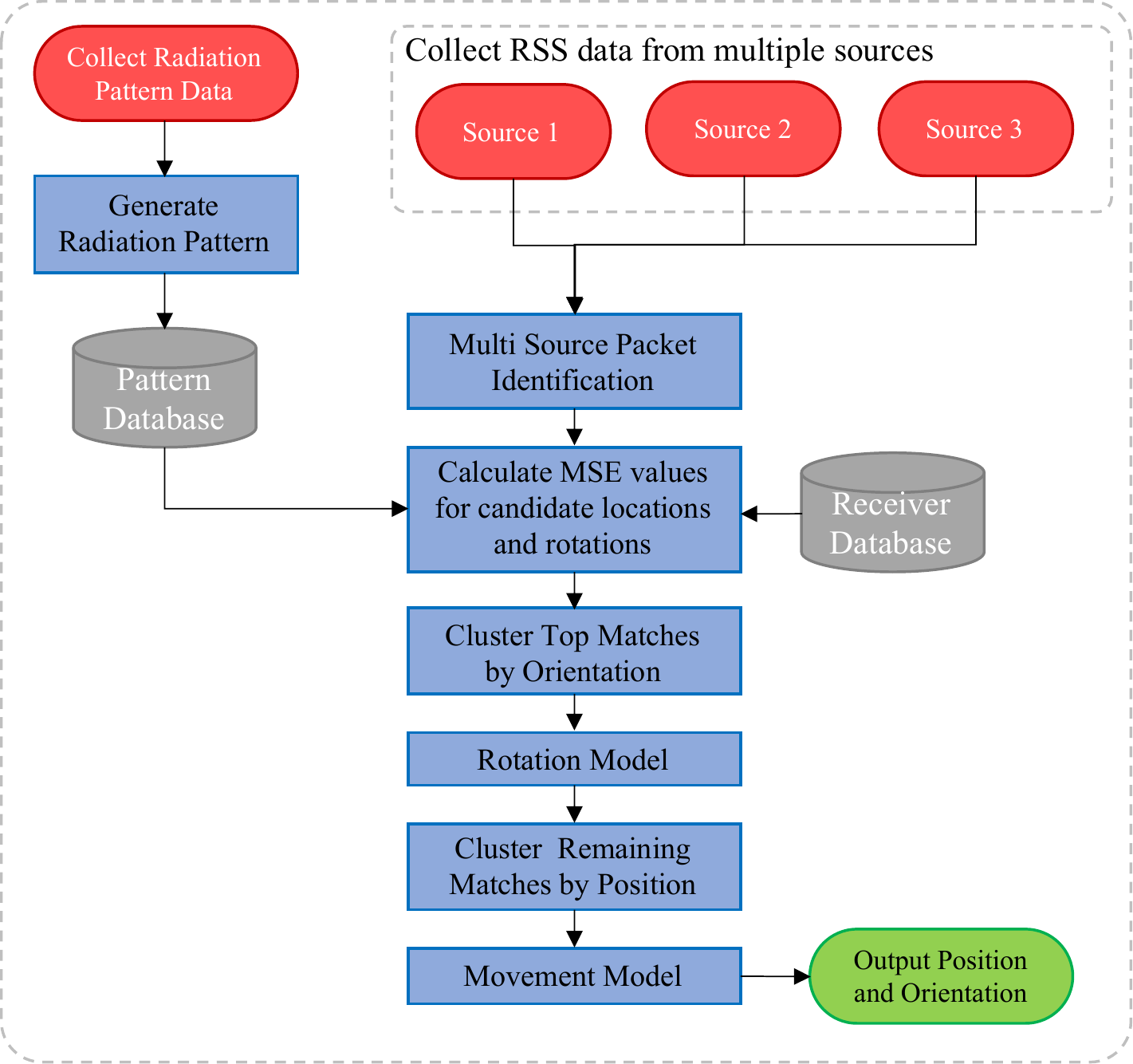}
	\caption[Enhanced Localisation Pipeline]{Enhanced localisation pipeline}
	\label{fig:cLocalisation-systemModel}
\end{figure}

We now present the design of our localisation pipeline in full. We first discuss the axis conventions used. We then discuss the individual steps of our pipeline. An overview of the different pipeline elements is shown in Figure~\ref{fig:cLocalisation-systemModel}.

\subsection{Axis Conventions}

This scheme relies on two separate coordinate systems, as the coordinate systems for device rotation and storing radiation patterns are different. We consider the environment a cartesian coordinate system for device rotation and movement to calculate the relative orientation between two devices. Radiation patterns are stored and accessed using a polar coordinate system.

\begin{figure*}[t]
   \centering
\begin{subfigure}{.5\textwidth}
  \centering
  \includegraphics[width=1\linewidth]{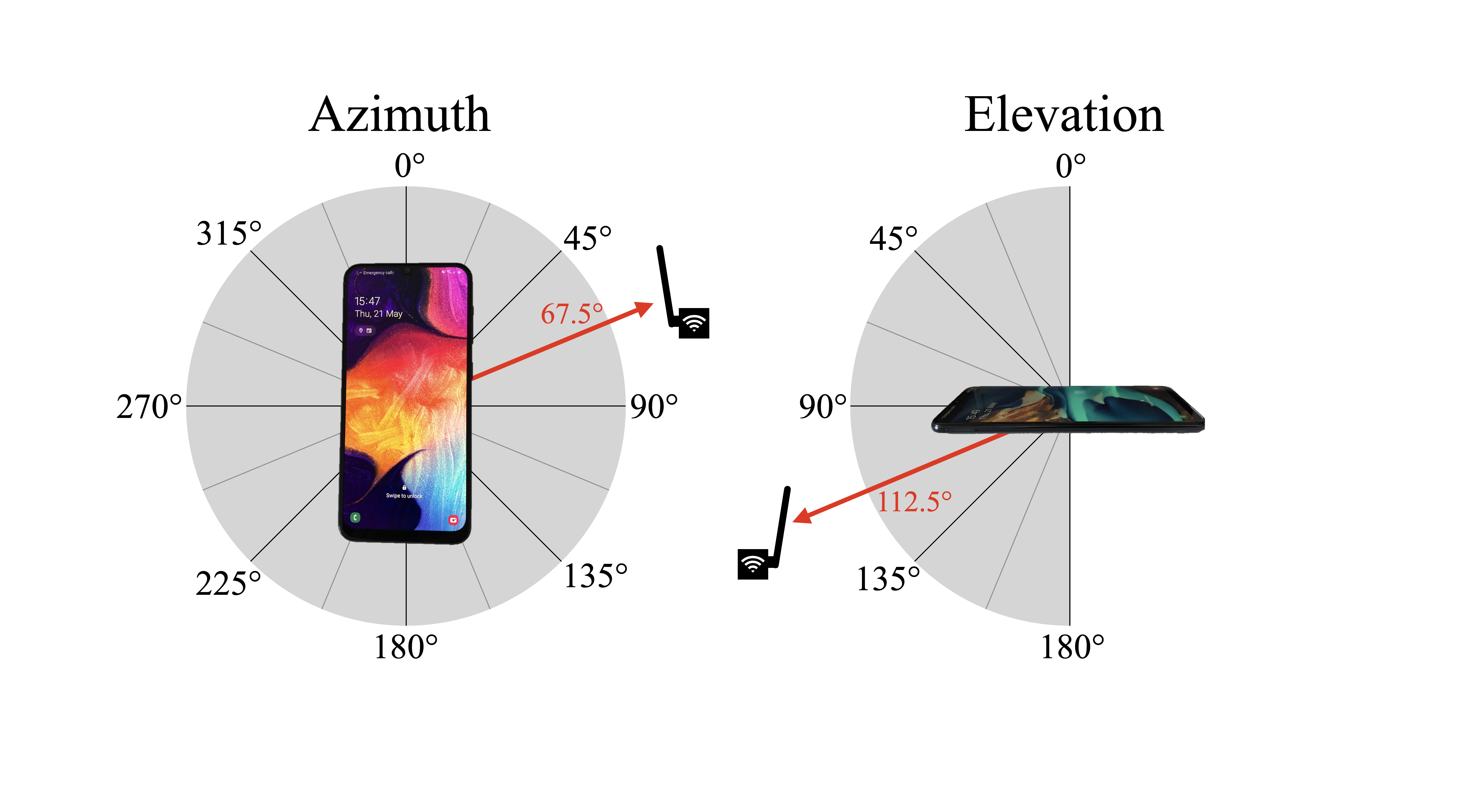}
  \caption{Pattern Measurements and Access}
  \label{fig:resolutionConventions}
\end{subfigure}%
\begin{subfigure}{.5\textwidth}
  \centering
  \includegraphics[width=1\linewidth]{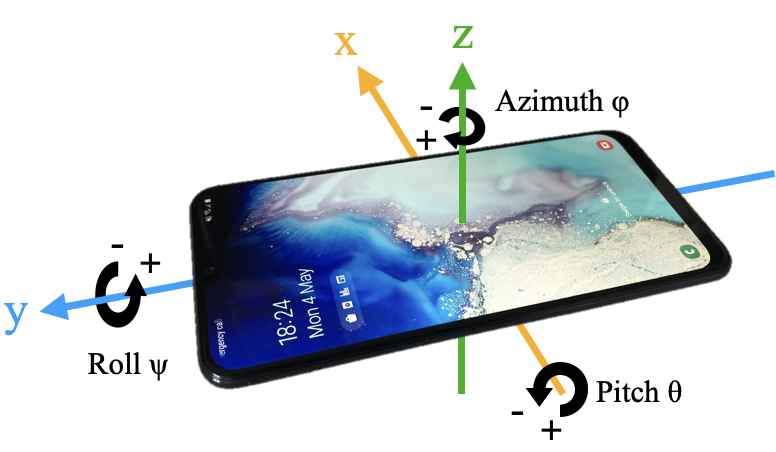}
  \caption{Device Rotation}
  \label{fig:rotationConventions}
\end{subfigure}%

\caption[Axis Conventions]{Axis conventions}
\label{fig:conventions}

\end{figure*}

We consider the resting state of the phone to be flat on a horizontal surface with the screen facing upwards and when the top of the phone is pointing in the direction of the positive y-axis of the environments reference frame, where x and y are horizontal and z is vertical. Azimuth ($\phi$) is rotation around the z-axis, pitch ($\theta$) is rotation around the x-axis, and roll ($\psi$) is rotation around the y-axis. The relative orientation of two devices must be resolved to 2 dimensions to access the pattern data structure as it uses a polar coordinate system. The dimensions are azimuth ($P_\phi$) and elevation ($P_\theta$). A visual representation of the axis conventions are shown in Figure~\ref{fig:conventions}(b) for device rotation and Figure~\ref{fig:conventions}(a) for polar coordinate resolution and pattern access.

\subsection{Pattern Enrolment}\label{sec:cLocalisation-patternEnrolment}

A radiation pattern must be collected for any device that one may wish to localise. This is performed by rotating the devices around two axes at set increments, e.g. 1 degree, at a fixed distance from a receiver. The receiver must be in the far-field region of the antenna to ensure the pattern is fully formed \cite{rahmat-samii1995,yaghjian1986}.
During collection, each RSS sample is saved along with the rotation values of the device at the time of transmission ($\textit{RSS},\phi,\theta$). 

As the receiver is static and the transmitter is rotating, it is necessary to invert the $\phi$ values because if, during collection, the device is rotated $45\degree$ clockwise the actual direction from the transmitter to the receiver is now $45\degree$ anti-clockwise. $\phi = 360 - motorAzimuth$ and $\theta = motorElevation$. With the following limits, complete pattern coverage is achieved with minimal duplicates: $ 0 \leq \phi < 360 $ and $ 0 \leq \theta < 180 $.

\begin{figure}[t]
	\centering
	\includegraphics[width=1\columnwidth]{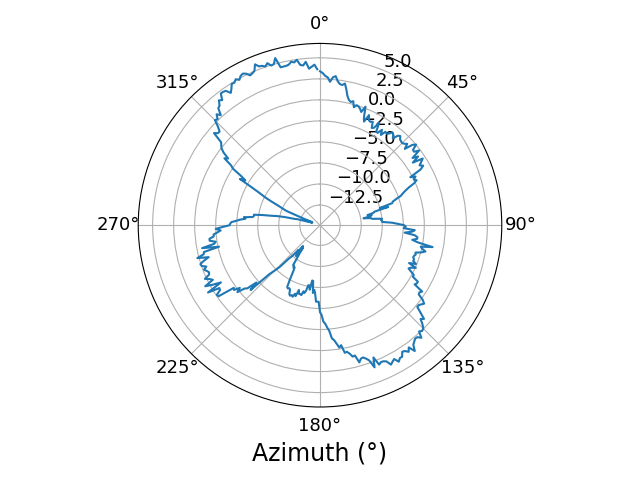}
	\caption[Samsung A50 Radiation Pattern]{Samsung A50 RSS (dBm) at approximately 5m}
	\label{fig:a50-pattern-raw}
\end{figure}

This list of tuples is then converted into a 2D array, which is a more efficient data structure for later access. An average is taken where there are multiple samples for a particular direction. The full pattern can now be accessed as $pattern[\phi][\theta]$. An example of a polar coordinate representation of this is shown in Figure~\ref{fig:a50-pattern-raw}.

\begin{figure*}[t]	
\begin{subfigure}{1.0\textwidth}
  \centering
  \includegraphics[width=1\linewidth]{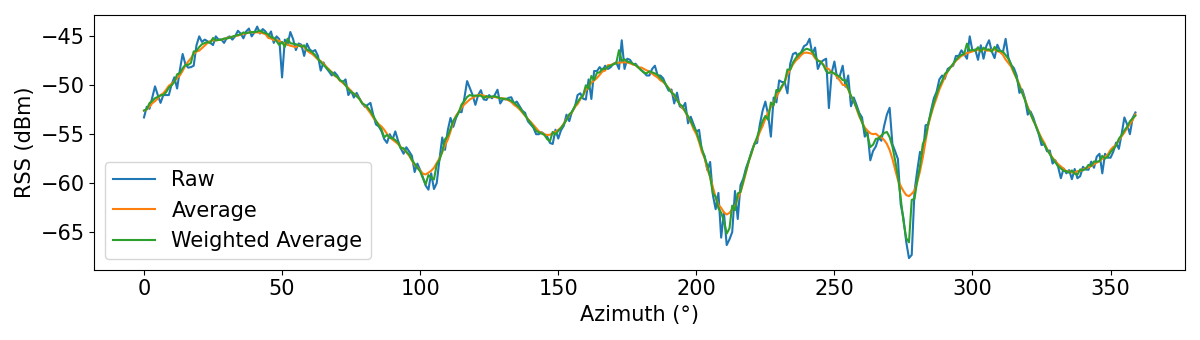}
  \caption{Incorrect method to smooth signal strength data.}
  \label{fig:bad-smoothing}
\end{subfigure}
\begin{subfigure}{1.0\textwidth}
  \centering
  \includegraphics[width=1\linewidth]{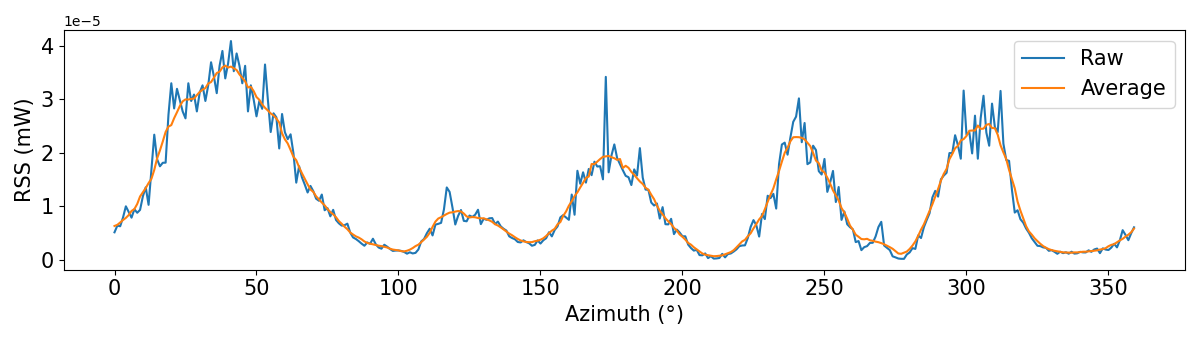}
  \caption{Correct method to smooth signal strength data.}
  \label{fig:good-smoothing}
\end{subfigure}
  \caption[Smoothing Raw Pattern Data]{Comparison of smoothing (a) log and (b) linear data. Both show the raw RSS data in blue, with an average being calculated for multiple measurements for the same azimuth value. Again for both, the orange line shows the method using the rolling average. On (a), the green line shows an unsuccessful attempt at weighting the different values depending on the distance from the centre of the rolling average window.}
  \label{fig:smoothing}
\end{figure*}

Even with calculating averages for each azimuth value, the pattern is not smooth. Without a smoothing function applied, severe location and orientation jitter may manifest in the final output. When displayed on a logarithmic scale, the troughs are sharper than the peaks. If the average of surrounding points is used to smooth logarithmic data, this will remove troughs while preserving peaks. The data is first converted from dBm to mW, as is shown in an example in Figure~\ref{fig:smoothing}, to ensure trough preservation. A similar process is performed by \cite{mwila2014}, although the rationale behind it is not explained. The values for each azimuth could then be smoothed by calculating a rolling average. However, the values on both sides of an azimuth measurement matter for the calculation. Therefore, we position the value to calculate at the centre of a sliding window and calculate the mean using values within 5$\degree$ in each direction. Figure~\ref{fig:smoothing}(a) shows the result of smoothing the data in dBm units, clearly demonstrating the failure to preserve the troughs. The green line shows the sliding window method but with a stronger weighting applied to the central values. This also shows inadequate smoothing when compared to Figure~\ref{fig:smoothing}(b).

Finally, the directivity must be calculated from the smoothed RSS data, and this becomes the pattern that is stored for later use.

\subsection{Packet Synchronisation}

We must know the power received by each receiver for each packet. When the data from multiple receivers is sent to the CS it must be processed to match each packet. COTS WiFi cards do not have an internal clock, so the CPU attaches timestamps. Therefore, time synchronisation cannot purely be relied on for packet identification. It is also important to consider that some packets may be lost due to collisions at a receiver or because a receiver is out of range.

Packets are matched from different receivers based on two criteria: 1) packet contents, a SHA256 hash of the encrypted packet. This does not guarantee uniqueness, but 2) looser time synchronisation can then be used. It is assumed that loose time synchronisation to some degree can be maintained (For example, 5ms in the case of using time synchronised Raspberry Pis).

A packet is uniquely identified if only one packet with a matching hash is found within the looser time tolerance. 
The output of this process is a list of packets with the signal strengths, timestamps, and receiver IDs.

\subsection{Likelihood Estimation}\label{sec:cLocalisation-dfe}

Potential candidate position and orientations are scored to find the most likely estimates for position and orientation. The candidate positions form a point grid in the possible space that $T$ could be located in. We use a linear-time search for simplicity as we only use a small number of receivers. For every candidate position and orientation, the difference from expected value is calculated for each packet received by computing the mean squared error (MSE). First, the expected values at the candidate position and orientation are calculated, and then the difference to the actual value collected is calculated. A site survey is conducted using a \textit{survey transmitter} with a known radiation pattern, and the expected values can be calculated by combining the site survey RSS measurements with the radiation pattern at different orientations.

\paragraph{Directivity from Pattern}

To access the pattern value in the direction of the receiver with a given orientation of the device, the vector defining the relative position of the receiver from the transmitter must be rotated using a rotation matrix ($M$) generated with candidate rotation values of the transmitter $T$ ($T_a$,$T_p$,$T_r$). $M = R P A$ where $R$, $P$, and $A$ are the rotation matrices for roll, pitch, and azimuth.

$M$ and the vector defining the relative position of the receiver from the transmitter gives the coordinates in 3D cartesian space that represent the final relative direction.
\begin{equation}
	\textit{coords} = M \cdot [r_x - T_x, r_y - T_y ,r_z - T_z]
\end{equation}
$coords$ is then converted from cartesian to polar coordinates to give us $relativeAzimuth$ and $relativeElevation$. The 2D \textit{pattern} array can then be accessed using these two values to fetch the directivity value for the transmitter ($D_{t}$).
\begin{equation}
	\label{eq:rssFromPattern}
	D_{t} = pattern[relativeAzimuth][relativeElevation]
\end{equation}

\paragraph{Site Survey}

The site survey is performed to measure the RSS values at many locations in the area where the system will operate. This will provide a list of positions that have RSS samples measured for all of the receivers in range. The transmission power of the survey transmitter and the directivity is removed from the measured RSS values to reduce the number of calculations needed later. This leaves us with a value corresponding to the path loss and effect of the receiver's directivity.
\begin{equation}
	M = R - P_{tx} - D_{st}
\end{equation}
where $R$ is the raw RSS measurement, $P_{tx}$ is the transmission power of the survey transmitter, and $D_{st}$ is the directivity of the survey transmitter in the direction of the corresponding receiver. This will leave the system with $\textit{candidatePositions} = [\{M_0,...,M_n\},...,\{M_0,...,M_n\}]$. The values in between the survey measurements are interpolated to allow for a more fine-grained coverage of the area than a site survey can achieve.

As the frequency can significantly impact the received signal strength \cite{christmann2010}, care also needs to be taken to ensure that the site survey is carried out using the same wireless channel or channels that will be used during the system's lifetime. This means that the wireless network must be fixed to a specific selection of frequencies once the site survey is complete, or a more extensive site survey covering the whole WiFi frequency range must be performed.

\paragraph{Expected RSS Value}

For each candidate position and orientation pair, the expected RSS value $\textit{rss}_\textit{expected}$ can be calculated for each receiver using:
\begin{equation}
	\textit{rss}_\textit{expected} = M_n + P_{tx} + D_t
\end{equation}
where $M_n$ is the previously calculated RSS value from the corresponding candidate position for that receiver, $P_{tx}$ is the transmission power, and $D_t$ is the directivity of the transmitter in the direction of the receiver based on the candidate position and orientation.

\paragraph{Difference From Expected}

For each candidate position and orientation, the difference from expected value is calculated using the mean squared error of the euclidean distance between the actual RSS measurement and the expected RSS value for each receiver.
\begin{equation}
	\label{eq:dfe}
	\textit{MSE} = \dfrac{1}{nReceivers} \sum_{i=1}^{nReceivers} { \left( R_{i_{(rss_{expected}})} - R_{i_{(rss_{actual}})} \right)^2}
\end{equation}

\subsection{Clustering Best Matches}

Ideally, the position and orientation with the lowest MSE would give the correct result. This is not always the case, it may be that many points far away from the actual location have a lower MSE value. This is particularly an issue when we consider the radiation pattern of the device, as this produces multiple regions or arcs with a strong match.

To account for this, the top $k$ MSE matches are selected as in existing k-nearest neighbour schemes. However, we select a much larger number to attempt to find multiple regions in our case, and in our examples and evaluation, we use $k$=100. The top $k$ are clustered for position and orientation using \textit{Density-Based Spatial Clustering of Applications with Noise} (DBSCAN) \cite{ester1996}. DBSCAN is ideal for this application because it can find non-linearly separable clusters, the number of clusters varies, and outlier points are expected. DBSCAN functions by examining neighbours of points. If a point is within a threshold distance ($\varepsilon$) of a threshold number of other points (\textit{minPts}) including itself, then it is categorised as a core point, and multiple core points can connect to become a cluster. There is also a category of points called non-core points that are within the threshold distance of a core point but do not reach the threshold number of neighbours. It is a member of the cluster but is an edge point and cannot connect up to other non-core points. All other points not categorised as core or non-core are excluded from clusters as outliers.

Position and orientation are clustered separately as there is not always a 1-to-1 match of clusters for position and orientation (i.e., it is possible to have scenarios where there are a greater or fewer number of position clusters than orientation clusters).

\paragraph{Orientation Clustering}
The top 100 orientation tuples are cluster using DBSCAN with parameters $\varepsilon$=2 and \textit{minPts}=10. For each axis of rotation, the same top 100 values are clustered for that axis and MSE value. The centroid is calculated using a weighted average to find the orientation of each cluster. The MSE values are normalised and inverted for the weights, so the lowest MSE value of the 100 top matches becomes 1, and the highest becomes 0.

The data can wrap around from 360 to 0 as there is rotation continuity. To form clusters across this boundary, we duplicate the data between -360 \& 0 and 360 \& 720. For every real cluster, there may now be up to 3. The clusters with a centroid greater or equal to 0 and less than 360 are considered.

\paragraph{Rotation Model}
A rotation model is now used to remove outliers based on rotation. The rotation model is applied to the device's orientation on each axis. Unlike \cite{hindmarch2016}, we are not considering the change in the direction of movement but the change in orientation of the device itself.

We assume that any device will be under the user's control, i.e. it is not dropped or thrown. This binds the movement of the device to the movement of the user. To constrain maximum rotation of a smartphone under the control of a human user, a maximum of 180$\degree$ of rotation in each axis over time $t$ seconds is applied. This is a tuneable parameter, and in our case, we use 500ms.

For the orientation to be valid, this must be within the bounds described above when compared to the previous valid orientation. If it is invalid in any axis, then the cluster is invalid, and all data points in that cluster are eliminated.

Using the model, after 500ms it is no longer possible to constrain rotation. The smaller the interval between packets, the greater the constraint on movement by the rotation model, and the more effective it will be at eliminating outliers. To ensure the security guarantees hold, there is a maximum limit on the time interval between valid positions, and this limit is set as $t/2$.

\paragraph{Position Clustering}
The remaining position tuples that were not declared as invalid by the rotation model are clustered using DBSCAN with parameters $\varepsilon$=0.5 and \textit{minPts}=10. Depending on the specific implementation, these parameters may need to be adjusted (e.g., larger $\varepsilon$ if the site survey interpolation points are sparse). Each cluster's position is the weighted centroid of the core points, using the same normalised weighing as before.  

A movement model is now applied to eliminate position clusters further away from the previous valid point than is feasible to move in the time between packets. Like the rotation parameter, this is tunable based on the domain. We use a maximum velocity of 10m/s as an absolute upper bound of human movement. Any clusters that would require a velocity of more than 10m/s are eliminated.

\paragraph{Final Scoring}
If multiple clusters remain after removing outliers, then a score for each cluster is calculated. The cluster score is the sum of the squared normalised and inverted MSE values using the same normalised values as weightings for the cluster centroid calculations. The cluster with the highest score is chosen. If there is not a 1-to-1 mapping of the clusters, then some extra steps need to be performed. Either the orientation or position must be selected as the primary scoring data. The cluster in the primary data with the highest score is found, and all other data points across the position and orientation data are eliminated if they are not in the highest scoring cluster. The weighted mean is then calculated for the non-primary data.

The pipeline outputs a stream of estimated positions ($x$,$y$,$z$), orientations ($\phi$,$\theta$,$\psi$), and timestamps for each valid location.

\subsection{Bounding Rotation}\label{sec:cLocalisation-bounding-rotation}

When using bounded movement, the packet interval plays an essential part in applying the limit. 
While the effectiveness of this limit may be reduced if the target is moving very slowly or the velocity limit is set too high, the maximum time between the samples does not need to be considered too carefully as the target is essentially on an infinite plane. As long as the target does not move beyond the velocity limit, then the movement is valid. However, once enough time has passed to allow for 180$\degree$ rotation, it is no longer possible to provide any bounds, as any orientation is now possible. Thus requiring a fairly high packet rate that cannot drop below a certain threshold without reducing accuracy. This is particularly important to consider in this case, as a smartphone can rotate very quickly, even under human control. In other fields where the transmitter would rotate slower under normal operating conditions, this would likely have less impact (e.g., automotive or aviation context).

\section{Evaluation} \label{sec:cLocalisation-experiments}

Like previous works on this topic \cite{nasipuri2002,mwila2014,zuo2019}, we use a combination of real measured data and simulated data for our evaluation. We collected radiation patterns from real devices and used simulations to evaluate this technique's potential increase in accuracy.

\subsection{The Significance of Smartphone Radiation Patterns}\label{sec:cLocalisation-signif-patterns}

The greater the variation in directivity across the sphere of the transmitter, the easier it will be to use RSS to measure orientation as it will be a larger factor in the overall RSS value measured.

A variety of smartphone patterns were measured for a single elevation slice to see this variation. Each device was placed on a platform and rotated $360\degree$ four times at a fixed distance of approximately 4.40m from a Raspberry Pi with an external WiFi card and antenna configured to sniff traffic. The device was set to transmit WiFi packets continuously. The device was rotated in $1\degree$ increments to have one or more packets transmitted at each degree increment for each rotation. The data collected were processed as described in Section~\ref{sec:cLocalisation-patternEnrolment}.

\begin{figure*}[t]
   \centering
\begin{subfigure}{.33\textwidth}
  \centering
  \includegraphics[width=1\linewidth]{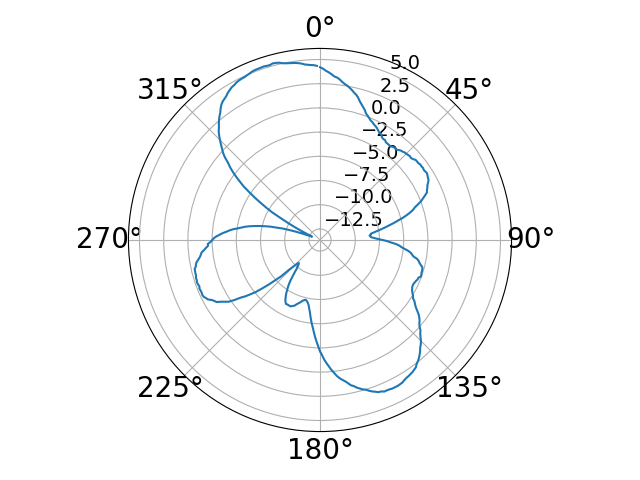}
  \caption{Samsung A50}
  \label{fig:sfigpattern1}
\end{subfigure}%
\begin{subfigure}{.33\textwidth}
  \centering
  \includegraphics[width=1\linewidth]{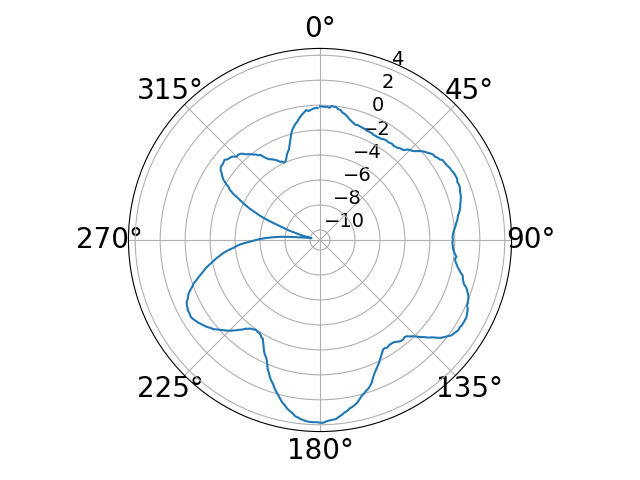}
  \caption{iPhone 6S}
  \label{fig:sfigpattern2}
\end{subfigure}%
\begin{subfigure}{.33\textwidth}
  \centering
  \includegraphics[width=1\linewidth]{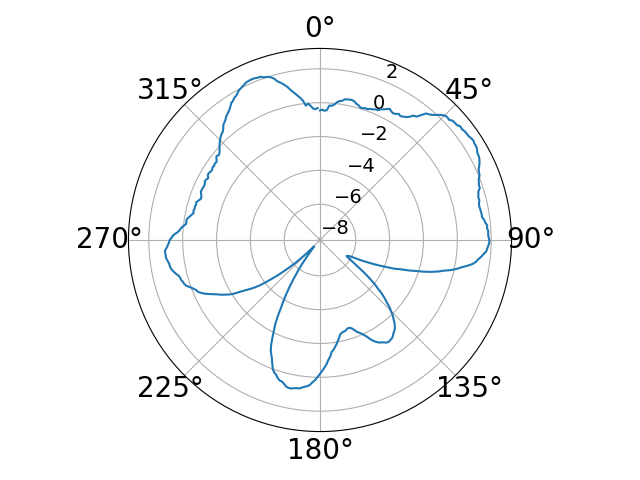}
  \caption{iPhone X}
  \label{fig:sfigpattern3}
\end{subfigure}
\begin{subfigure}{.33\textwidth}
  \centering
  \includegraphics[width=1\linewidth]{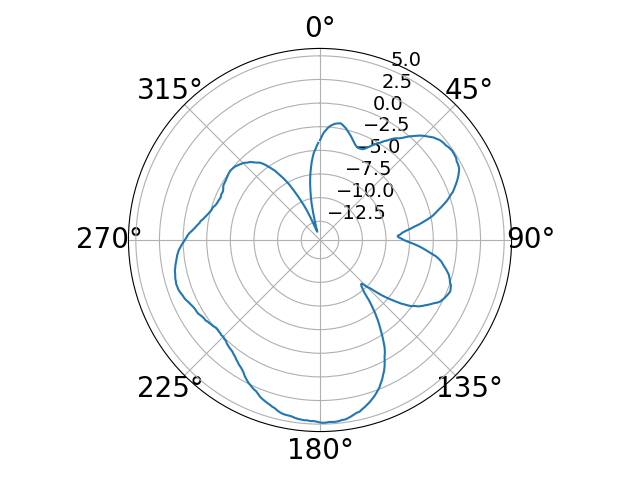}
  \caption{iPhone 5S}
  \label{fig:sfigpattern3}
\end{subfigure}%
\begin{subfigure}{.33\textwidth}
  \centering
  \includegraphics[width=1\linewidth]{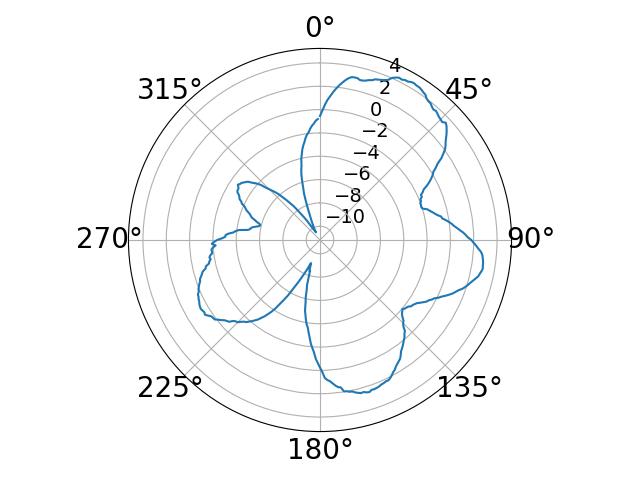}
  \caption{Moto E3}
  \label{fig:sfigpattern3}
\end{subfigure}%

\caption[Example Radiation Patterns]{Plots of radiation pattern slices (in dBi) of WiFi antenna at 90$\degree$ elevation. 
}
\label{fig:cLocalisation-patterns}

\end{figure*}

These showed significant variation in directivity depending on the direction and differences across a range of devices. For example, at 90$\degree$ elevation, the Samsung A50 had a maximum directivity of 5.06dBi and a minimum of $-12.29$dBi. Figure~\ref{fig:cLocalisation-patterns} shows the result of 5 different devices.

\subsection{2D Proof of Concept Simulations}

\begin{figure*}[t]
  \centering
  \includegraphics[width=\textwidth]{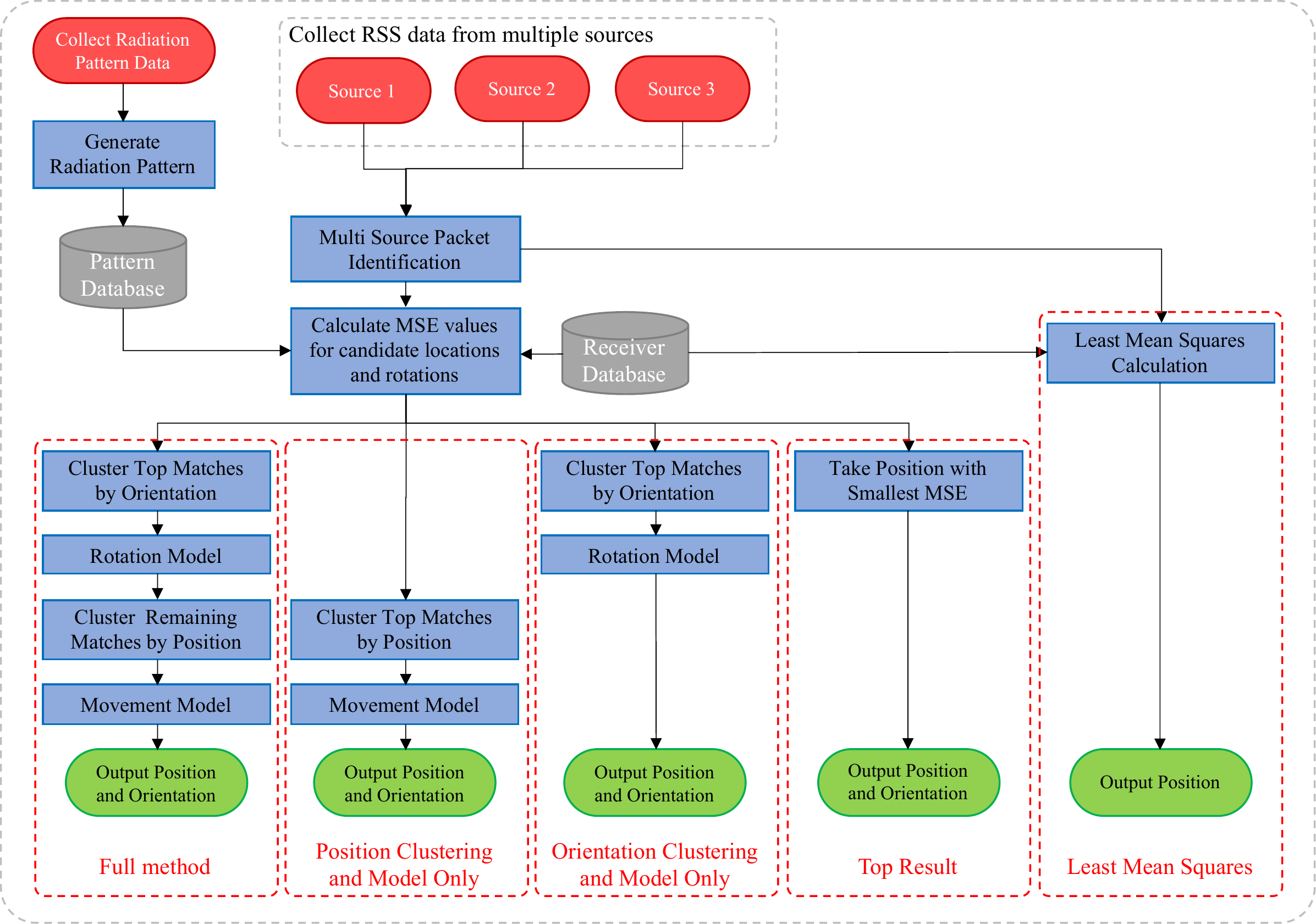}
  \caption[Testing Model]{The evaluation was performed by testing the separate elements contained in the red dashed boxes.
  }
  \label{fig:cLocalisation-testingSystemModel}
\end{figure*}

To test the robustness of the pipeline against fast-fading noise under different conditions, we used simulated datasets like the most closely related work \cite{mwila2014,zuo2019}. We passed these simulated datasets through a modified version of the localisation pipeline, shown in Figure~\ref{fig:cLocalisation-testingSystemModel}, to evaluate the separate components individually and as a whole. This was to determine what level of improvement was possible compared to the case of not considering radiation patterns and measure the contribution of each step of the localisation pipeline to ensure no component does not add anything to the process. The modified version of the pipeline uses five comparison methods: 1) \textit{full method}, the full enhanced localisation pipeline; 2) \textit{location clustering and model only}, where the steps of orientation clustering and rotation model are removed; 3) \textit{orientation clustering and model only}, where the location clustering and movement model components are removed; 4) \textit{top result}, the naive approach where we take the top matching position and orientation as the estimate; and finally 5) \textit{least mean squares} (LMS), where the estimate for location is determined using log distance model and a non-linear least mean squares optimiser (LMS) without considering the radiation pattern, and this does not produce an orientation estimate. We use the latter two as baselines for comparison, the LMS method as the baseline for location and the top result method as the baseline for orientation.

The environment was modelled as a 20m x 20m environment with between 3 and 6 receivers. The transmitter moved incrementally between a pair of positions and rotated while the receivers were fixed in position and orientation. The site survey RSS values were calculated using a log distance path loss model based on the position, orientation and radiation patterns of the receivers. The survey used 1m x 1m grid squares with interpolated steps of 0.1m. Gaussian noise with various standard deviations $\sigma$ was added as the $X_{\sigma}$ component of the path loss model. The simulated datasets were passed through the pipeline, and then the output location and orientation results were compared to the known position and orientation.

In most simulations, the transmitter moved from an initial starting position to an end position in a straight line at a fixed velocity, and linear rotation was applied to the device. Packets were transmitted every 5ms, and the noise was added. The positions, velocities, and rotation rates were varied for 10 different simulations.

A variety of $\sigma$ values for noise were used to reflect a range of different real environments. To select realistic $\sigma$ values, RSS samples were taken from a static device over a period of time in real environments. These environments included transmission with a direct line of sight and through multiple walls/rooms. These values had standard deviations in the range of 1.5 to 2.6 dBm. Thus, $\sigma$ values 0.5, 1.0, 1.5, 2.0, 2.5, and 3.0 dBm were used for the simulations.

Additionally, after the initial simulations were complete, the following parameters were modified to reflect different scenarios: 1) the number of receivers was adjusted; 2) A longer packet interval was used; and 3) non-linear periodic motion was also used to reflect human movement better.

The simulated data was generated using the parameters discussed, and this data was passed through a modified version of the enhanced localisation pipeline (shown in Figure~\ref{fig:cLocalisation-testingSystemModel}) implemented using a combination of Java for the initial MSE calculations and Python for the clustering stages.

It is hard to directly compare the results of our simulations to the simulations from existing works due to the significant differences as laid out in the introduction to this paper. But we chose simulation parameters (e.g., size of environment, interpolation step, default pattern resolution, path loss exponent, etc.) that gave us results for mean position error in a similar range as previous work \cite{mwila2014} and we compare the relevant parts of our results to \cite{zuo2019}.

\subsubsection{Simulation Results}

We now present the results of our simulations by first examining the accuracy of the position and orientation from the simulations in general. Then we delve deeper into other factors including, transmission interval, number of receivers, rotation outside of the bounds of the rotation model, and initial synchronisation.

The results are presented as the absolute error values and relative values in comparison to the baselines of the LMS and top result methods. The absolute values are important in the respect that they show what degree of accuracy is feasible, but the relative values are more interesting when it comes to comparing to the baseline methods.

\paragraph{Location}

\begin{figure*}[t]
  \centering
  \includegraphics[width=0.83\textwidth]{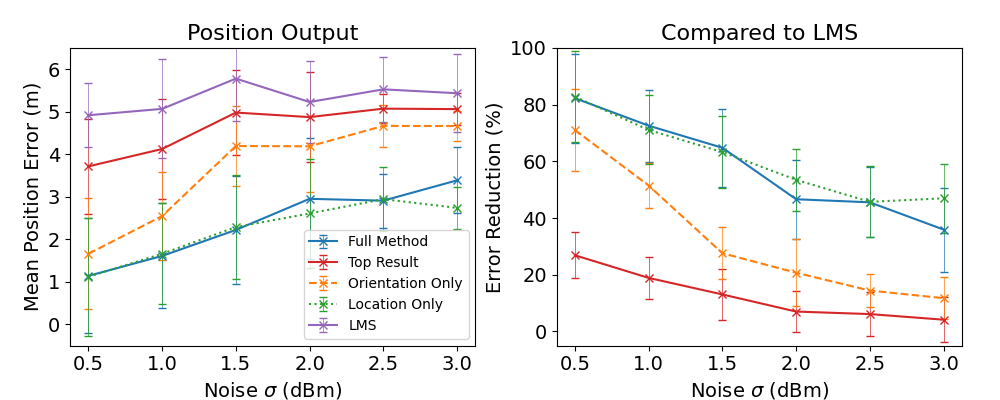}
  \caption[Simulation Results of Position Error]{Result of simulations showing the absolute mean position error and reduction in mean position error versus the least mean squares optimiser and the top position result for different levels of noise. 4 receivers were used for the simulation. Error bars show the 95\% confidence intervals from 10 simulation runs.
  }
  \label{fig:cLocalisation-4ReceiverResults-location}
\end{figure*}

In terms of mean location error, as shown in Figure~\ref{fig:cLocalisation-4ReceiverResults-location}, the results of the simulations demonstrate that the full localisation pipeline (full method) outperformed the LMS optimiser and top result baselines. Relative to the LMS optimiser, the full method showed a 82.28\% and 35.94\% reduction in the mean position error for the lowest and highest noise levels, respectively. The relative improvement over both of the baseline methods reduces as the level of noise increases, indicating that the full method is more susceptible to noise than either of the two baseline methods.

As expected, each of the four methods comprised of pipeline components outperformed the LMS method. When comparing the four methods, taking the top result consistently produced the least accurate results. The full method and location clustering method perform very similarly. Finally, the orientation clustering and model only method slightly underperformed the full and location clustering methods.

Our results demonstrate that the full method performs better or similarly compared to any of the sub-components individually.

\paragraph{Orientation}
Switching focus to the output orientation values of the methods, we cannot compare the result of orientation to the LMS optimiser as this has no orientation output, so we solely use the top result as a baseline. 

\begin{figure*}[t]
  \centering
  \includegraphics[width=0.83\textwidth]{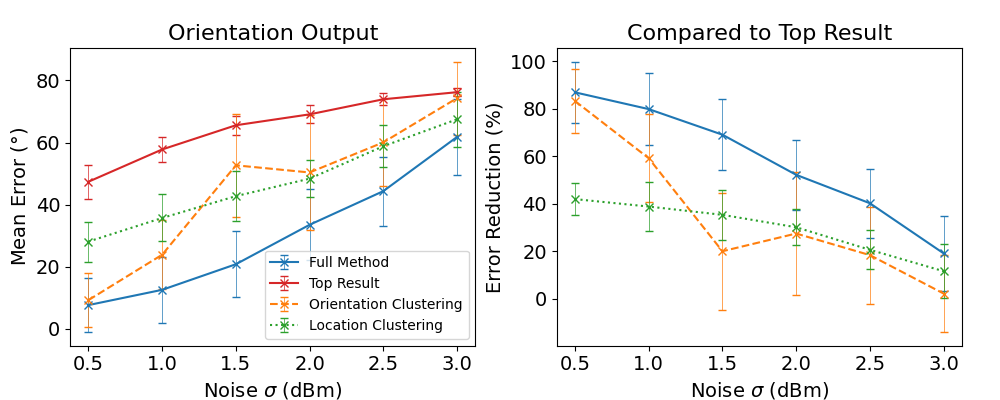}
  \caption[Simulation Results of Orientation Error]{Result of simulations showing the reduction in mean orientation error and the reduction in error versus the top result method. 4 receivers were used for the simulation. Error bars show the 95\% confidence intervals from 10 simulation runs.  
  }
  \label{fig:cLocalisation-4ReceiverResults-orientation}
\end{figure*}

The results are shown in Figure~\ref{fig:cLocalisation-4ReceiverResults-orientation}. The left plot shows the mean orientation error for the different pipeline components, and the right shows the relative error reduction compared to taking the top result. These showed good estimates of orientation. At the lowest noise level, a mean orientation error of 7.6$\degree$ was achieved and at the highest noise level, the mean orientation error was 61.8$\degree$. Unlike with position error, the results show that the full method outperforms each sub-component individually in terms of mean orientation error, showing that the full method is required to perform best. The orientation clustering and location clustering methods produce similar results at the higher levels of noise tested.

The results demonstrate that combining all elements is superior in estimating the orientation over any individual component. When compared to \cite{zuo2019} our method can achieve similar levels of orientation accuracy with far fewer nodes, just 4 receivers versus their 10-50 nodes, although they are able to achieve higher accuracy overall due to their very different system model with up to 200 nodes.

\paragraph{Number of Receivers}
	
	The first factor we examine more closely is the number of receivers. As the number of receivers increases in most localisation schemes, the number of constraints on the position of the device increase, and so in general, the accuracy will increase. Our proposed scheme is no different when considering the mean position and orientation error.
	
\begin{figure*}[p]
  \centering
  \includegraphics[width=\textwidth]{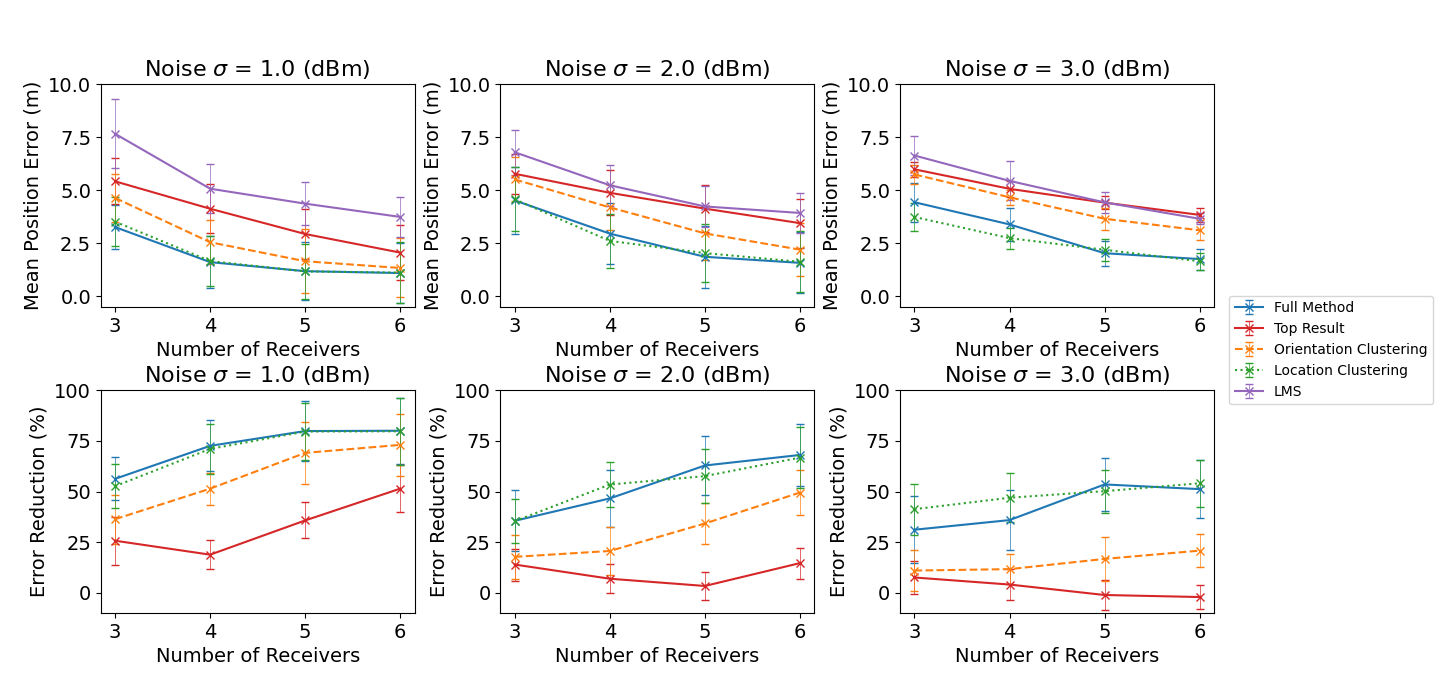}
  \caption[Simulation Results of Position Error, Number Receivers]{Result of simulations showing the reduction in absolute mean position error  and mean position error versus the least mean squares optimiser with varying numbers of receivers. Error bars show the 95\% confidence intervals from 10 simulation runs.
  }
  \label{fig:cLocalisation-nReceiverResults-location}
\end{figure*}

\begin{figure*}[p]
  \centering
  \includegraphics[width=\textwidth]{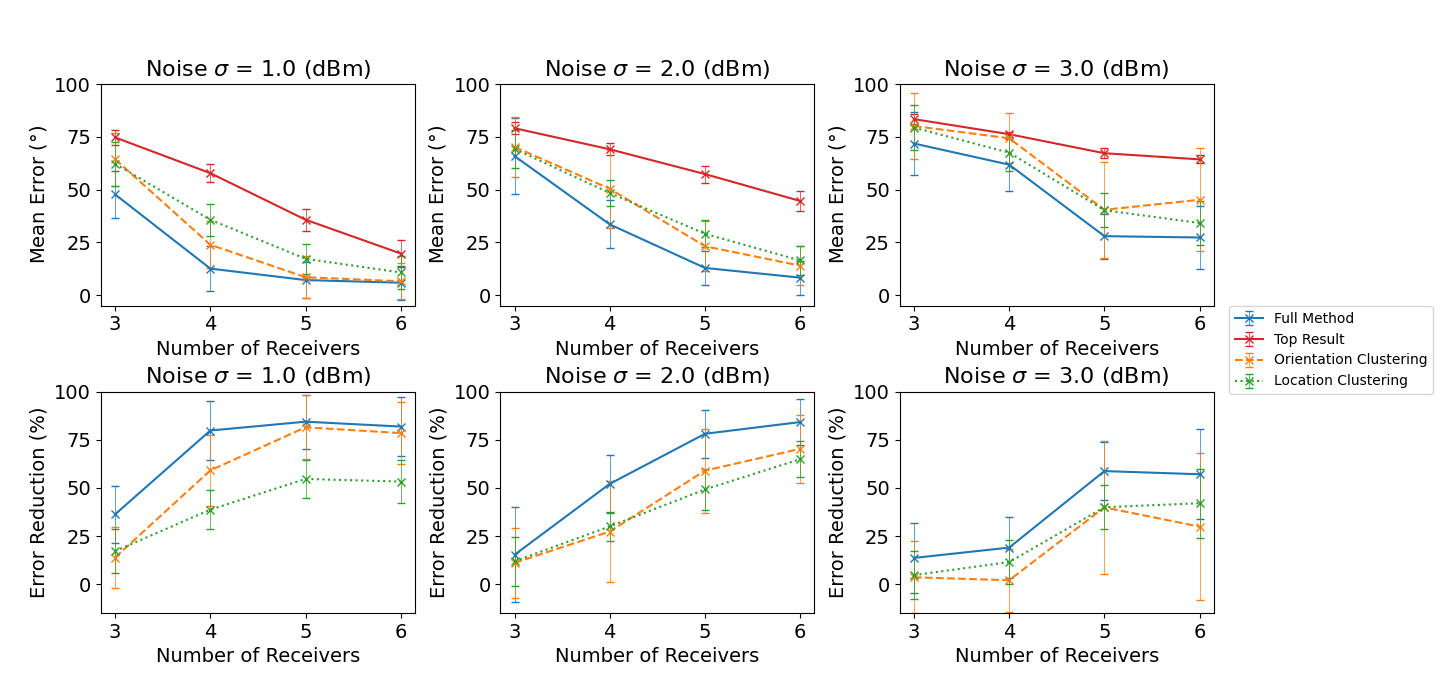}
  \caption[Simulation Results of Orientation Error, Number Receivers]{Result of simulations showing the reduction in absolute mean orientation error and mean orientation error versus the top result. Error bars show the 95\% confidence intervals from 10 simulation runs. 
  }
  \label{fig:cLocalisation-nReceiverResults-orientation}
\end{figure*}

	We examine the relationship of the number of receivers and mean error in terms of absolute error and relative error. The relevant data are shown in Figures~\ref{fig:cLocalisation-nReceiverResults-location}~and~\ref{fig:cLocalisation-nReceiverResults-orientation}. First focusing on position error, results shown in Figure~\ref{fig:cLocalisation-nReceiverResults-location}, the absolute values show that the error is reduced for all of the methods tested as the number of receivers increases. As we would expect, there is a greater improvement in accuracy for datasets with more noise. The benefits of adding more receivers trail off for the clustering methods but not for the top result, and this intuitively makes sense. For clustering, the centroid of the cluster is taken. This may not be symmetrical around the true position. Adding more receivers will likely pull the estimated position closer to the true position when just taking the top match.
	
	When compared to the baseline LMS optimiser, the results show that the error reduction improves with more receivers for the full method, location clustering only, and orientation clustering only. Meanwhile, the top result method results are more mixed with no significant relative improvement for the higher noise levels. 
	
	Secondly, focusing on the orientation error, Figure~\ref{fig:cLocalisation-nReceiverResults-orientation} shows similar results to that of the position error. In general, there is a decrease in mean orientation error as the number of receivers is increased and the top result method performs the worst.

\paragraph{Transmission Interval}

\begin{figure*}[p]
  \centering
  \includegraphics[width=\textwidth]{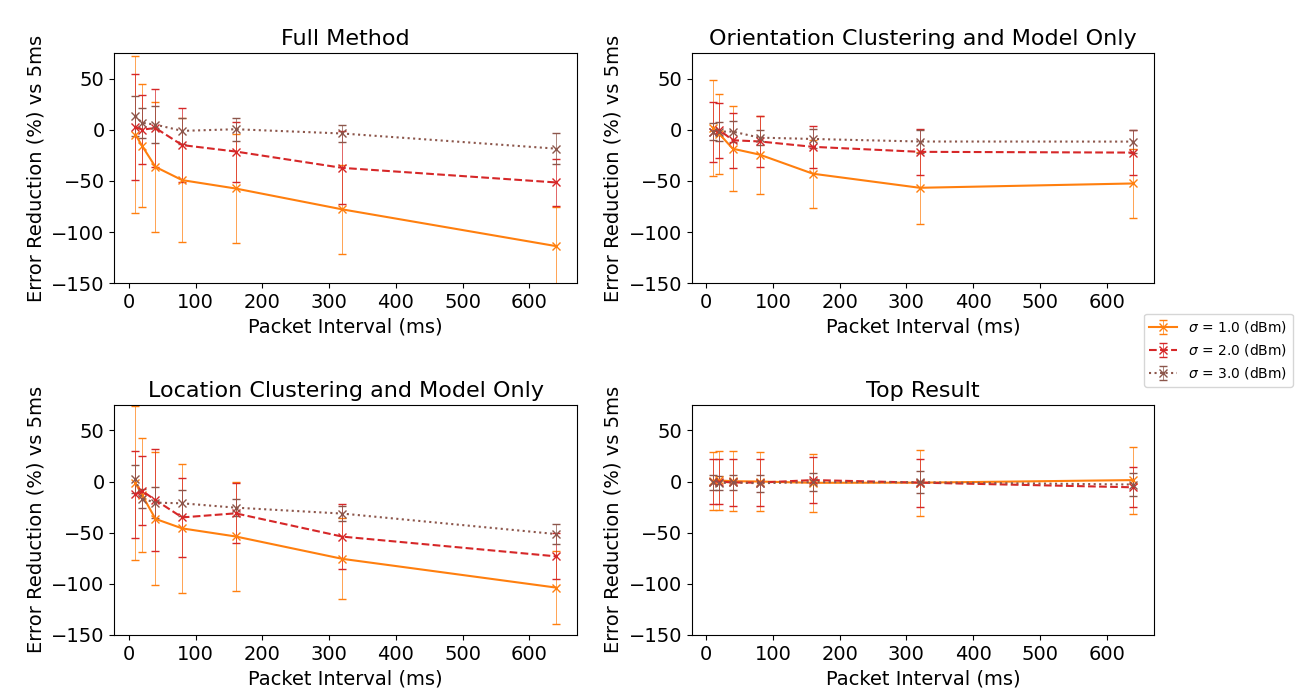}
  \caption[Simulation Results of Position Error, Packet Interval]{Result of simulations showing the difference in mean position error with different packet intervals versus 5ms packet intervals. Error bars show the 95\% confidence intervals from 10 simulation runs.
  }
  \label{fig:cLocalisation-packetIntervalResults-location}
\end{figure*}

\begin{figure*}[p]
  \centering
  \includegraphics[width=\textwidth]{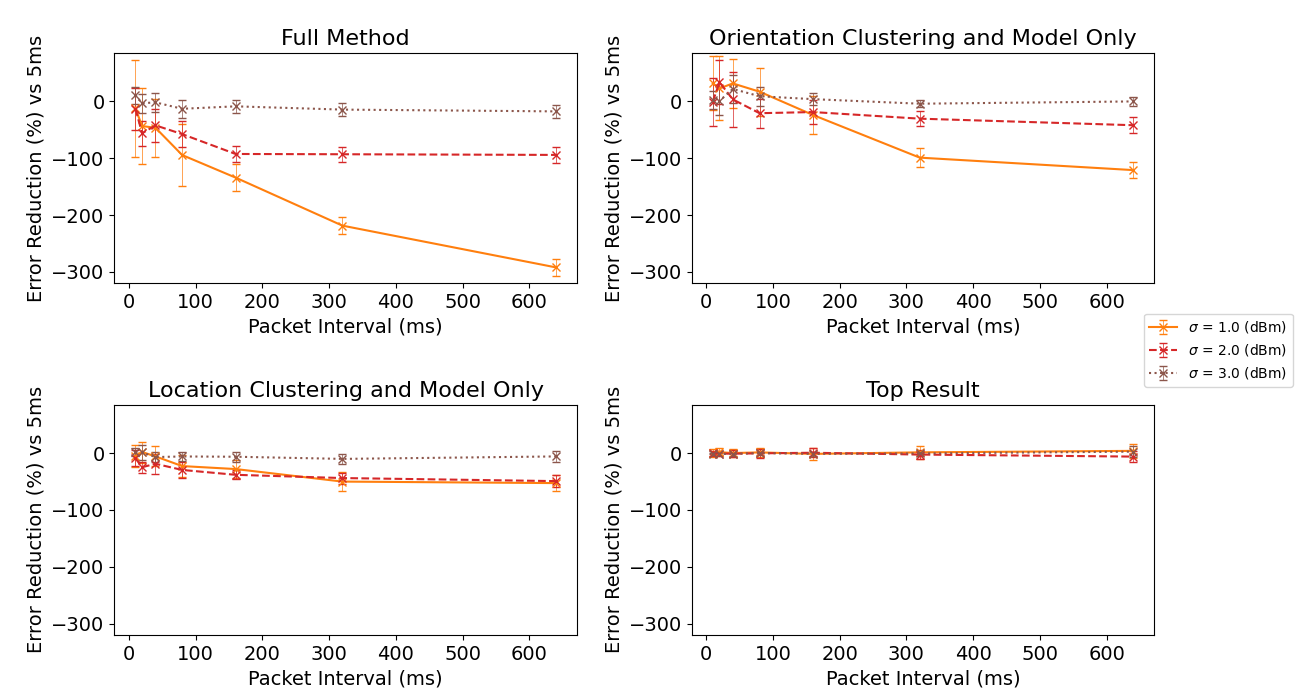}
  \caption[Simulation Results of Orientation Error, Packet Interval]{Result of simulations showing the difference in mean orientation error with different packet intervals versus 5ms packet intervals. Error bars show the 95\% confidence intervals from 10 simulation runs.
  }
  \label{fig:cLocalisation-packetIntervalResults-orientation}
\end{figure*}

The transmission interval (i.e., the time interval between packets) is very important because as the time between packets increases, the constraint on the rotation is reduced, as previously discussed in Section~\ref{sec:cLocalisation-bounding-rotation}. Figures~\ref{fig:cLocalisation-packetIntervalResults-location}~and~\ref{fig:cLocalisation-packetIntervalResults-orientation} show the relationship between packet interval and error relative when compared to a 5ms packet interval in terms of position and orientation respectively. The results for location demonstrate a general trend of increasing error as the packet interval is increased for all different pipeline components and methods except for the top result method---which has no packet time calculation so should not be affected by the packet interval. 

When looking at the results for orientation accuracy, the behaviour is very similar for the location clustering and top result methods. However, the results show a more considerable drop-off in accuracy for the orientation clustering method and a fairly severe drop-off in accuracy for the full method for the lowest noise level simulated.

\paragraph{Periodic Rotation}
	The periodic non-linear rotation simulation---designed to more accurately reflect human movement---also performed similarly to the linear rotation simulation with corresponding $\sigma$ noise dataset.

\paragraph{Rotation Outside Bounds}
As expected, when the actual rotation is outside the bounds of the rotation model, the accuracy of both the position and orientation is reduced significantly. In this case, the pipeline essentially defaults to purely relying on the movement model to eliminate outliers at a less frequent rate.

\subsection{Performance}

The search of the candidate locations grid requires a brute force search. The simplest approach has a complexity of $\mathcal{O}(N_l \cdot N_o \cdot N_r)$ where $N_l$ is the number of candidate locations, $N_o$ is the number of candidate orientations, and $N_r$ is the number of receivers. As with any grid location search, the performance can be significantly improved once the first packet calculations are complete and the movement bounds are applied. Every subsequent packet only needs to be checked for candidate orientations and positions within the rotation model tolerance and movement model tolerance. 

Additionally, to improve performance, the expected RSS value for candidate locations and orientations calculations can be pre-computed as this is the most computationally expensive part of the pipeline. The actual RSS values can then be compared to each candidate without matrix multiplications. Also, each candidate comparison is independent and can be parallelised. Other domain-specific optimisations can be made, for example, not including inaccessible locations in the candidate grid.

\section{Security Analysis}\label{sec:cLocalisation-spoof}

The type of attack of greatest concern for localisation systems is, in general, a spoofing attack. This is where an attacker falsifies the measurements to make it appear as if they are in a different location than their true location. 
The problem with radiation patterns is that they present a way for an attacker to vary the ratios of power transmitted in a particular set of directions by changing the device's orientation. 
Attacks on RSS localisation schemes exploiting radiation patterns give the attacker more scope to find a matching position, giving them an advantage over an attacker that is restricted to an attacker model using an isotropic antenna.
An attacker can precompute a transmission power, position, and orientation that would result in signal strength measurements that allow them to spoof another location. 
In this analysis, we use \textit{location} to refer to the combined values of position and orientation, and we consider position and orientation to be equally important.

To attack the system to spoof an incorrect path, the attacker Mallory must carry out the following tasks for each waypoint location in the path:

Mallory must first calculate the expected RSS values for each receiver as if she was transmitting from the spoofed position $A_{x,y}$ and orientation $A_{\theta}$ with the pattern $P_a$ that the system believes Mallory is using. This is possible as Mallory has complete knowledge of the scheme and data in the databases. The calculated RSS measurements are $m = [Rss_0,...,Rss_n]$ and she must aim to match them from another location.

With the known expected RSS values $m$, from Mallory's current location plus any movement within the allowed bounds, Mallory must direct the correct part of her radiation pattern at each receiver so that the receiver measures the expected RSS values. As she has a fixed pattern, she cannot individually target each receiver, so the packet sent will be heard by any receiver within range simultaneously (time difference caused by the speed of light negligible in this case). Therefore, Mallory must find a position $B_{x,y}$ and orientation $B_\theta$ to satisfy: $(A_{x,y}, A_\theta) \Rightarrow m \land (B_{x,y}, B_\theta) \Rightarrow m \land dis(A_{x,y},B_{x,y}) > d$ while using her actual pattern $P_b$. This must be performed for each of the $k$ waypoints.

\paragraph{Same Pattern Attack}
We first consider that Mallory's pattern is the same as the pattern that the system believes she is using (i.e., $P_a = P_b$). 

We examine the probability of finding matches to determine how likely this situation is. We ran simulations using a similar setup and parameters to our evaluation simulations to determine the probability of finding matches. The area $\mathcal{A}$ was initialised with random receiver locations, and 100 random locations to spoof were selected. $\mathcal{A}$ was then searched for other possible locations that provided a strong match to each of the locations to spoof (i.e., close enough that they would appear in the top 100 best locations by MSE). We used 4 receivers and a distance threshold $d$ of 5. 

The probability of finding at least one match for a specific spoofed location from 20 runs of the simulations was 0.522 with a 95\% confidence interval of 0.040. This result is not unexpected, given that the radiation pattern gives the attacker a great deal of flexibility to find matches. 
However, to transmit at the calculated attack locations, the attacker must be able to move without exceeding the movement and rotation limits as specified by the adversary model. To test this, we kept track of all the possible locations that could be used to spoof a straight path through the environment. We then attempted to find a valid path for the movement and rotation model from the possible locations. We allowed the attacker to selectively not transmit when she could not find a valid location up to the $t/2$ limit mentioned earlier. From this simulation, we found that even small movements in the order of 10s of centimetres would require much larger changes in location for the attacker. With the movement and rotation of the attacker set to 1/10th of the movement limits, in out of 20 simulations the attack failed in less than 100 packets for all of them. Therefore, a sustained waypoint spoofing attack is difficult to achieve as the movement and rotation of the attacker device must consistently exceed the limits. 

It is important to note that in this case, as Mallory is within the system's area of operation $\mathcal{A}$, if she does find another matching location, both locations will appear as a strong match if she is using the pattern that is expected by the system.
Therefore, she cannot guarantee that the system will choose the spoofed path with a strong probability, making the attack even more difficult for her.

\paragraph{Different Pattern Attack}

We next consider the second case, Mallory's pattern is different from the pattern that the system believes that she is using (i.e., $P_a \neq P_b$). Unlike the previous case, this should not typically produce a strong match at Mallory's actual position.

We took two different patterns from the selection that we measured and ran the same simulations as we did for the same pattern attack. The probability of finding at least one match for a specific spoofed location was 0.547 with a 95\% confidence interval of 0.039. Therefore, an attacker does not have any significant advantage or disadvantage when using a different pattern compared to using the same pattern. When we considered the movement and rotation constraints, we found that movement and rotation exceeding the limits was required. Therefore, using a different pattern for a waypoint attack is also infeasible.

\section{Related Work}\label{sec:related}

Unlike in the area of acoustic localisation---where orientation detection has been achieved \cite{kim2011,sanchez-hevia2017}---orientation detection using time flight measurements with speed of light communication and COTS devices is much more difficult because of the timing precision required. This is why considering radiation patterns is essential for usable orientation detection.

	There are many possible ways to categorise and group existing localisation schemes that attempt to account for the impact of antenna or radiation patterns. The most appropriate categories to consider for this paper are 1) `simple' patterns and 2) `complex' patterns. In some existing work, the model of directionality is very simple. The pattern may be modelled as a single Gaussian-shaped lobe, a single cone, or a simple symmetric shape. The significant advantage this provides is that it is relatively easy to optimise algorithms to search for location matches or account for the error introduced by the pattern. They can also produce position estimates with very little ambiguity, as there are very few parts of the pattern with matching directivity. The disadvantage is that these are not a perfect match to the actual directional properties of the antenna. It is important to note that the inaccuracy varies significantly depending on the type of antenna, so in some situations, it is appropriate to use this `simple' pattern model. Methods that can be better characterised as using `complex' patterns, on the other hand, better match some patterns' real-world properties---this is particularly true for the work conducted in this paper. We will first cover existing work that uses a range of differently shaped 'simple' patterns. Finally, we will discuss work that uses 'complex' shaped patterns.

\paragraph{Elliptical Patterns}

A domain in which we have seen the consideration of radiation patterns is in cellular networks where Zhou et al. \cite{Zhou2005} use an elliptical propagation model rather than a circular one as a result of the directionality. This is an improvement over conventional localisation schemes, but it only considers the pattern of the base station.

\paragraph{Gaussian Patterns}

Zuo et al. \cite{zuo2019} attempt to solve the limitations of isotropic assumptions for localisation. The goal is to localise multiple directional transmitters with unknown transmission power and orientation using many receivers. They present three algorithms: an alternating projection (AP) algorithm, an expectation maximisation (EM) like algorithm, and a particle swarm optimisation (PSO) algorithm. These vary in complexity and accuracy, the PSO algorithm's most accurate but has the trade-off of having the slowest runtime. The other two are less accurate but run significantly faster. They make a number of assumptions that are substantially different from our work. Firstly, they assume that the directional sources have a Gaussian-shaped pattern with a fixed beam width. This is different from our work as we allow for any arbitrarily shaped pattern. Secondly, they assume that the receiving sensors are isotropic. Meanwhile, we assume any wireless device, including the receivers, can have any arbitrarily shaped patterns, which adds to the quantity of information we can use.

Li et al. \cite{li2020} use a combination of multiple antennas with known radiation patterns to calculate the distance and angle of a device from a single access point equipped with multiple antennas. They minimise the need for a complete site survey by using a path-loss model, the radiation pattern of the multiple antennas, and differential RSS. This has some similarities to other techniques that we will mention shortly. To remove outliers, they also apply a further data fusion step relying on dead reckoning using inertial navigation.

\paragraph{Single Beam Patterns}

The influence of radiation patterns on localisation has been explored in the deployment of sensor networks. These often use a single beam model of the antenna pattern, such as a narrow cone projecting from the device.

A scheme proposed by Nasipuri \cite{nasipuri2002} uses a small number of coordinated fixed position rotating beacons with a narrow beam that rotates at a fixed angular speed. The beacon orientations are each at a specific offset and are separately identifiable. Nodes can then use timing information to calculate the angle to each beacon and therefore find their own location.
Farman et al. \cite{farman2012} propose a method of cooperative localisation for sensor networks in which there are several rounds of adjusting the direction of a directional antenna and the transmission range of an anchor to locate sensor nodes. Gautman et al. \cite{gautam2019} present a technique to determine the location of a node in polar coordinates from an anchor with a directional antenna that can rotate around two axes.

Several schemes rely on mobile anchor nodes with directional antennas. In these schemes, the position of the anchor node is known using GPS or another positioning system. The location is broadcast as the node moves through the environment with static sensor nodes randomly positioned. These schemes aim to determine the position of the sensor nodes.
	Ou \cite{ou2011} takes the approach of using a 4 directional antennas on the anchor node pointing in 4 directions (positive $x$, negative $x$, positive $y$, negative $y$ axes). On the other hand, Chang et al. \cite{chang2018} using a single directional antenna but move the mobile node in a specific pattern. While they use different methods for the location calculation, they both rely on the $x$ coordinate and $y$ coordinate being calculated at different stages of movement. They both require the assumption that the antenna pattern has a main lobe and that it can be approximately represented as a cone with a fixed angle $\theta$, which does not necessarily accurately reflect radiation patterns for other types of wireless devices.

Another approach is to use rotating antennas or multiple directional antennas on an anchor to estimate the angle of arrival (AoA). Jiang et al. \cite{jiang2012} use this approach employing multiple such anchor nodes, rotating nodes and nodes with two antennas placed perpendicularly to each other. The single antenna rotating node is used to create a quadratic and a linear function during an initial gathering stage. These functions are then used to estimate the AoA between 0$\degree$ and 90$\degree$ using the static node with two perpendicularly mounted antennas. Using the AoA from two nodes enables localisation. Alternatively, an array of antennas can be used to in-effect remove the need for the antenna to rotate. Work by Passafiume et al. \cite{passafiume2017} combines this with dual-frequency antennas to additionally exploit the difference in pattern shape.
In the moving vehicle domain, ARREST \cite{ghosh2020a} is an autonomous RSS positioning and tracking system. The system employs a single rotating direction antenna onboard a moving vehicle to track and follow an RF-emitting beacon. It is, therefore, able to determine the direction of the tracked object from itself. In this example, the shape is not too important. The key criteria is that there is one particularly strong lobe. Our method does not rely on non-static receivers and uses the inherent directionality of all types of antennas.

\paragraph{Complex Patterns}

We now discuss related work using more complex patterns or arbitrarily shaped patterns which are a more accurate model of radiation patterns for devices like smartphones.

The impact of more complex radiation pattern shapes on localisation has received some attention. Coca \cite{coca2013} explored the impact of radiation patterns on range-based localisation schemes on wireless sensor network nodes. The authors proposed using additional onboard compasses and calibration certificates for Wireless Sensor Network nodes to correct for the pattern. Like the previously mentioned data fusion approaches, this approach would require access to the data provided by the onboard compass, which would not necessarily be available to the network infrastructure. We approach this from an infrastructure perspective, so we do not have access to onboard sensors. Mwila et al. \cite{mwila2014} developed a model for using radiation patterns in localisation by adding antenna error, caused by radiation patterns, as a component of the path loss model. They presented a Gauss-Newton optimisation method to improve on localisation schemes that do not consider radiation patterns. Through simulations, the authors found a 59.23\% improvement in average localisation error by considering radiation patterns over not considering them. Again, this was dependent on knowing the orientation of the nodes.

In scenarios where there are many sensor nodes, it is possible to estimate many of the properties of a transmitter. Bolea et al. \cite{bolea2011} used data from a large number of sensor nodes to remotely estimate certain features of the environment and transmitter, including transmission power, path loss exponent for the propagation model, antenna orientation, and the radiation pattern of the transmitter. Like many WSN approaches, this is dependent on having a large number of receivers, which is not appropriate for our application.

\section{Conclusion} \label{sec:cLocalisation-conclusion}

In this paper, we have presented a novel localisation scheme that uses radiation patterns of devices to filter out improbable changes in device orientation to provide an accurate estimate of the device's orientation and position. We presented the design our an enhanced localisation pipeline. We used simulations to show that the pipeline can eliminate incorrect regions of position and orientation caused by the additional ambiguity added to the localisation process by radiation patterns, even with realistic noise levels. The scheme achieved mean orientation errors between 7.6$\degree$ and 61.8$\degree$ across the range of RSS noise tested with just four receivers. A security analysis was performed that showed a path spoofing attack on the scheme is infeasible.



%\bibliographystyle{IEEEtran}
%\begin{thebibliography}{1}
%\end{thebibliography}

\bibliography{MyLibrary}
\bibliographystyle{IEEEtran}

\end{document}